\documentclass[12pt]{iopart}
\usepackage{psfrag}

\usepackage[T1]{fontenc} 
\usepackage[cp1250]{inputenc}
\usepackage{graphicx} 
\usepackage[english]{babel} 
\usepackage{color}

\newcommand{\bq}{\begin{equation}} 
\newcommand{\eq}{\end{equation}}
\newcommand{\ba}{\begin{eqnarray}} 
\newcommand{\ea}{\end{eqnarray}}
\newcommand{\degree}{$^\circ$}
\newcommand{\um}{$\mu$m}

\renewcommand{\vec}[1]{\underline{#1}}

\providecommand{\black}[1]{\textcolor{black}{#1}}

\begin{document}

\bibliographystyle{unsrt}

\topical[Statistical physics of bacterial growth]{Bacterial growth: a statistical physicist's guide}

\author{Rosalind J Allen and Bart{\l}omiej Waclaw}
\address{School of Physics and Astronomy, The University of Edinburgh, James Clerk Maxwell Building,
Peter Guthrie Tait Road, Edinburgh EH9 3FD, UK.}

\ead{rosalind.allen@.ed.ac.uk}

\begin{abstract}
Bacterial growth presents many beautiful phenomena that pose new theoretical challenges to statistical physicists, and are also amenable to laboratory experimentation. This review provides some of the essential biological background, discusses recent  applications of statistical physics in this field, and highlights the potential for future research. \\
\end{abstract}

\maketitle
\newpage


\section{Introduction}
Consider the following scenario: a small number  of pathogenic bacteria (perhaps 10-100) enter the human body and cause an infection. An antibiotic is prescribed to fight off the infection. Assuming that the bacterial infection is initially sensitive to the antibiotic, what are the chances of curing the infection, and how likely {\black{is it} that the infection eventually becomes resistant to the antibiotic? Given the increasing global health issue posed by antibiotic resistant infections, this is an important and timely problem \cite{organization_who_2014,organization_who_2001,us_antibiotic_2013,uk_uk_2013,oneill_2014}. Clearly, understanding the growth of bacterial infections and the potential for emergence and spread of antibiotic resistance within them requires collaboration between scientists from  different disciplines. Do statistical physicists have a role to play here? Furthermore, thinking more broadly, could our understanding of other processes mediated by bacterial growth, such as global biogeochemical cycles \cite{ob:fenchel}, human gut health \cite{flint_review}, and wastewater treatment \cite{ob:sludge}, also profit from a statistical physics-like approach?

Statistical physicists find inspiration in systems where complex macroscopic behaviour arises from a simple set of underlying microscopic dynamical rules. Living systems obviously belong to this class, and statistical physics has a long history of applications to biological problems. Examples include {\black{determination of mutation rates by analysis of mutant number statistics \cite{luria_mutations_1943,lea_distribution_1949}}}, the totally asymmetric exclusion process  \cite{blythe_review}, which was originally proposed as a model for protein production from messenger RNA in biological cells \cite{mcdonald}, {\black{models for noise in gene regulation \cite{elowitz_2002,thattai_noise}}}, lattice models of growing populations \cite{eden,court,waclaw_spatial_2015}, {\black{models for collective flocking and swarming behaviour \cite{vicsek,toner,gregoire}}} and non-equilibrium phase transitions in populations of self-propelled "swimmers" {\black{\cite{fily,buttinoni,marchetti,stenhammar,cates_review2}}}.

In this review we argue that the dynamics of growing bacterial populations provides another class of systems to which the methods of statistical physics can naturally be applied. To briefly illustrate this, we notice that the above example of the growth  of an antibiotic-resistant infection involves stochastic phenomena on scales ranging from macroscopic to molecular. Specifically:\\

\noindent {\it Macroscopic level: population expansion in space.} In many real-world scenarios, including infections, bacterial populations spread in space (e.g. through an infected tissue). This process could be modelled using the Fisher-Kolmogorov equation,
\bq
	\frac{\partial n(x,t)}{\partial t} = D \frac{\partial^2 n(x,t)}{\partial x^2} + r n(x,t)(1-n(x,t)/K),
\eq
where $n(x,t)$ is the population density of bacteria in space and time, $r$ is the maximal growth rate, $D$ is a diffusion constant that accounts for bacterial motility and $K$ represents a maximal population density. Section \ref{sec:hetgrowth} will discuss applications of this equation, and other approaches, to bacterial populations growing in heterogeneous environments.\\

\noindent {\it Microscopic level: bacterial replication processes.} The growth dynamics of a population containing  a mixture of antibiotic-sensitive and antibiotic-resistant bacterial cells, which replicate stochastically with rates $r_S$ and $r_R$, could be described by the following simple master equation:
{\setlength{\mathindent}{0.5cm}
\bq
	\frac{dP(S,R)}{dt} = r_R (R-1)P(S,R-1) + r_S (S-1)P(S-1,R) - (r_R+r_S)P(S,R),
\eq}
\noindent
where $S$ and $R$ are the numbers of sensitive and resistant cells. This type of equation can be solved using methods developed by statistical physicists to study processes such as random walks, birth-death processes and coalescence processes, as we shall discuss in Section \ref{sec:homgrowth}.\\

\noindent {\it Molecular level: gene expression.} Resistance to an antibiotic can be caused by genetic changes in the bacterial DNA (mutations), or by changes in how the bacterium expresses its genes. To express a gene, the DNA sequence is first transcribed, or copied, into an mRNA transcript molecule, which is then translated into protein (see section \ref{sec:biol}). To model this process we can use a set of Langevin equations:
\ba
		dx/dt = k_x -\gamma_x x + \eta_x(t), \\
		dX/dt = k_X x -\gamma_X X + \eta_X(t),
\ea
where $x$ and $X$ are the concentration of the mRNA and protein respectively, $k_x$ and $k_X$ are the transcription and translation rates, $\gamma_x$ and $\gamma_X$ are decay rates, and $\eta_x(t)$ and $\eta_X(t)$ represent Gaussian noise. These equations are similar to those  encountered in other statistical physics problems.\\

In this review, we discuss how statistical physics models can be applied to problems in bacterial population dynamics. The purpose of this review is to encourage interest in these problems, and to provide some of the basic biological background that is needed to appreciate the field. Physics and biology are of course very different in their language, philosophy, background and culture, and  full immersion into the world of bacteria comes with considerable challenges. 
Nevertheless, we hope to show here that bacterial population growth phenomena can provide considerable inspiration for the development of new and interesting statistical physics models. 

All the models that are discussed in this review are idealized and abstract descriptions of complex biological processes. It is often necessary to formulate coarse-grained models for biological systems, because many of the underlying details (e.g. interactions or rate constants) are simply not known. The most difficult aspect of the problem may not be how to solve the model, but how to formulate it so that it is coarse-grained enough to provide useful insight, but takes account of the essential biology, allowing it to give useful predictions. In many cases, the "right" physical or mathematical model of a biological system depends on the question that one is trying to answer.

We begin by introducing the reader to some basic microbiology, and to some interesting collective phenomena exhibited by bacterial populations. We do not aim at a comprehensive introduction, but rather we try to provide just enough information to follow the topics discussed later in the review --  more detailed background material  is available in excellent textbooks \cite{growth_bacteria_book,schaechter_microbe._2006,ob:Brock}. The remainder of this review is devoted to a more detailed description of bacterial growth phenomena which  present interesting challenges for statistical physicists, and examples of how statistical physics has been applied to these problems. This is divided into two parts, which cover growth phenomena in well-mixed systems and in spatially heterogeneous systems, respectively.  Finally, we present our own perspectives on the potential of this field, and on the relationship between statistical physics and microbiology. For lack of space, we do not consider in this review the fascinating and important topic of bacterial evolution, where statistical physicists have also made major contributions (for example to understanding the structure of fitness landscapes). Here, we refer the reader to the excellent review by De Visser and Krug \cite{de_visser_empirical_2014}.

\section{Background}

In this section we give a very brief introduction to the basics of bacteria and their growth. We also introduce the reader to several different types of stochastic collective behaviours that are exhibited by bacterial populations. Table \ref{tab:numbers} contains useful numbers relating to some of the topics that we discuss in the text.

\begin{table}
\centering
\begin{tabular}{p{8cm}|p{8cm}}
\hline
 {\bf Parameter} & {\bf Value} \\
\hline
 size & typical width 1\um, typical length 2-5\um\\
\hline 
 exponential growth rate & maximum: $\sim$2h$^{-1}$, sub-optimal conditions: 0.3--1.5 h$^{-1}$ \\ 
\hline 
 minimum doubling time & $\sim$ 20 min \\
\hline 
 elongation rate & $0.1-0.2\mu$m/min (on rich medium) \\
\hline 
 maximum density & $\sim$ 1--5 $\times 10^9$ cells per ml (LB medium, stationary phase) \\
\hline 
 mutation rate & $\sim 2\times 10^{-10}$ per bp per replication \cite{lee_rate_2012}\\
\hline 
 glucose molecules consumed to make 1 cell & $\sim 1.8 \times 10^{10}$ \cite{growth_bacteria_book,kk_egli_1998}\\
\hline 
 weight & 280 fg per cell \cite{neidhardt_book}\\
\hline 
 protein molecules per cell & $2.35\times 10^6$ (1850 distinct protein molecules) \cite{neidhardt_book} \\
\hline 
 mRNA molecules per cell & 1380 \cite{neidhardt_book} \\
\hline 
 genome size & $4.5\times 10^6$ bp \\
\hline 
 genome copy number & 1 (slow growth) to 8 (fast growth) \cite{nordstrom_copy-number_2006} \\
\hline 
 abundance of RNA polymerase & $\sim$ 1\% of total protein mass \cite{bremer_dennis} \\
\hline 
 abundance of ribosomes (growth rate dependent) &  $\sim$20-40\% of total mass \cite{neidhardt_book}; $\sim$ 7000-70,000 per cell\\
\hline 
 DNA replication rate & 580 - 1,190 bp/s \cite{bremer_dennis}\\
\hline 
 mRNA elongation rate (transcription) &  39-56 nucleotides/s \cite{bremer_dennis}\\
\hline 
 peptide elongation rate (translation) & 13-22 amino acids/s \cite{bremer_dennis}\\
\hline 
 intracellular concentration of ATP (growth in glucose-fed chemostat) & 9.6mM \cite{bennett} \\
\hline 
 intracellular concentration of a typical metabolite & 0.1-100mM \cite{bennett} \\
\hline 
 total intracellular metabolite concentration & $\sim$300mM \cite{bennett} \\
\hline 
 plasmid size & $\sim$2-500kbp \\
\hline 
 plasmid copy number & $\sim$1-200 per cell \\
\hline 
 minimal inhibitory concentration, & \\
 - ampicillin (inhibits cell wall synthesis) & $\sim 8\mu$g/ml (LB medium) \\ 
 - rifampicin (RNA synthesis inhibitor) & $\sim 3\mu$g/ml  (LB medium)\\ 
 - ciprofloxacin (DNA gyrase inhibitor) & $\sim 20$ng/ml (LB medium) \\  
\hline
\end{tabular}
\caption{\label{tab:numbers}Useful numbers for modelling bacterial growth and evolution. All data refers to the bacterium {\it E. coli}. If no reference is given, the values come from in-house experiments for the MG1655 strain of {\it E. coli}. We also refer the reader to bionumbers \cite{ref:bionumbers} - an excellent source of biology-related numbers.}
\end{table}

\subsection{Basic microbiology for statistical physicists}\label{sec:biol}

From a statistical physicist's point of view, a bacterium can be viewed as a microscopic particle, or cell, which is bounded by a pair of membranes with a stiff wall in between them (specifically, this is the case for a large class of bacteria that are known as Gram negatives; Gram positive bacteria have a thicker wall and lack the outer membrane). The interior of the bacterial cell contains a ``soup'' of DNA (encoding the bacterial genome), RNA, proteins, and other molecules (Fig. \ref{fig:1}). The materials that make up the bacterium are generically referred to as ``biomass''. Bacterial cells come in different shapes: from rods, to spheres, to spirals  (Fig. \ref{fig:1}), and sizes: from $\sim$100nm to $\sim$100$\mu$m. {\em Escherichia coli}, the  ``workhorse" of the microbiology lab, is a spherocylindrical Gram negative bacterium whose cells are $\sim 0.8-1\mu$m in diameter and $\sim 2-4\mu$m in length \cite{neidhardt_book,growth_bacteria_book}. 

\begin{figure}
\centering
\includegraphics[width=0.7\linewidth]{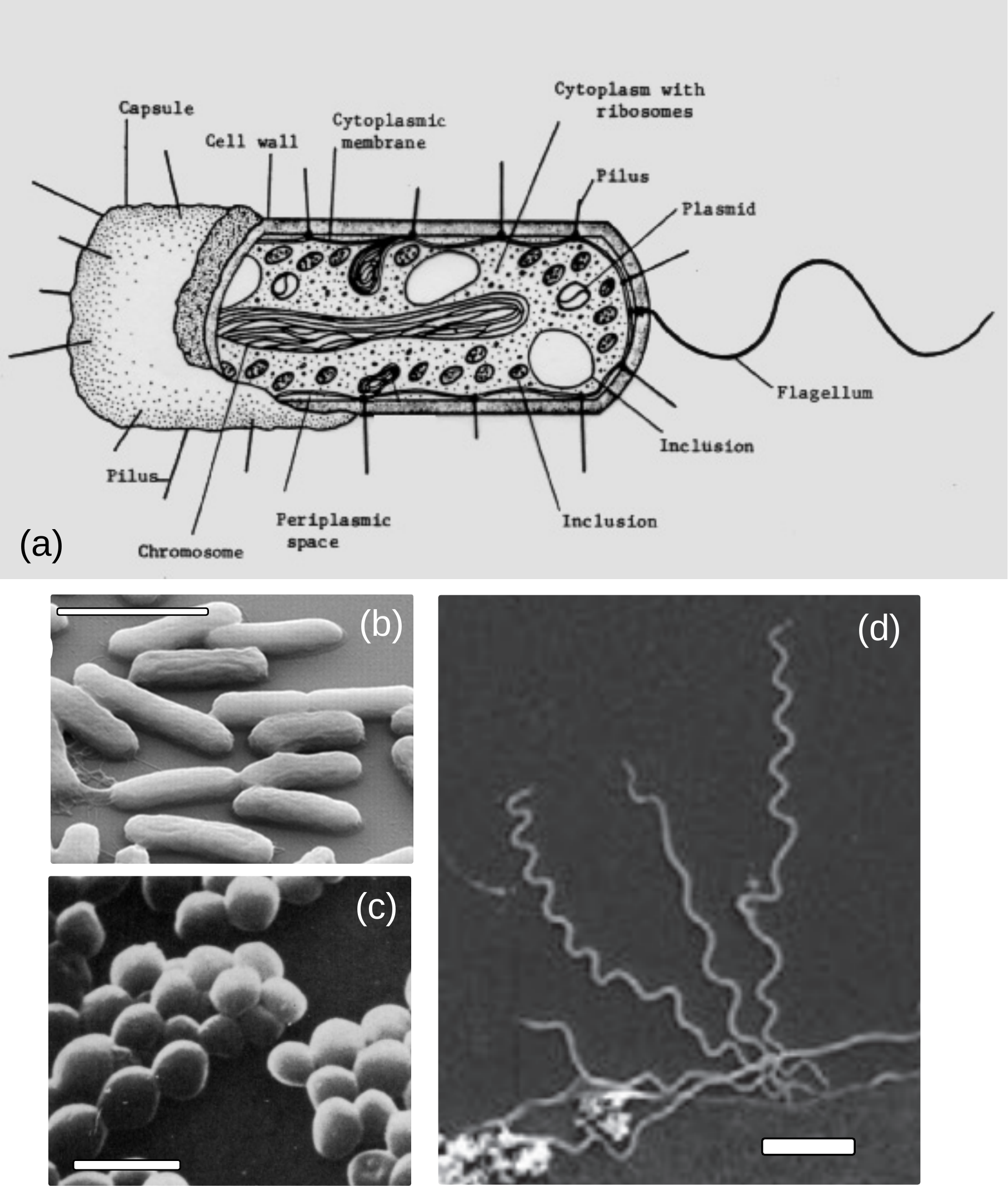}
\caption{(A) Schematic illustration of the structure of a typical Gram-negative bacterium, reproduced from {\em Todar's Online Textbook of Bacteriology}. (B) Scanning electron micrograph of {\em Escherichia coli} cells on a silicon surface, reproduced from Ref. \cite{hartmann}. (C) Scanning electron micrograph of {\em Staphylococcus aureus} cells, reproduced from Ref. \cite{greenwood_staph}. (D) Scanning electron micrograph of {\em Treponema pallidum} cells (the causative agent of syphilis), adhering to a human brain epithelial cell, reproduced from Ref. \cite{wu}. In panels B-D, the scale bars represent 2$\mu$m.}
\label{fig:1}
\end{figure}

Bacterial growth consists of the conversion of chemical nutrients into biomass. Nutrients enter the bacterium through pores in its membrane and undergo a series of chemical transformations, converting them into new cellular components; these chemical transformations are collectively known as metabolism \cite{growth_bacteria_book,schaechter_microbe._2006}. The increase in biomass is accompanied by an increase in cell size and by replication of the bacterial DNA, possibly with some errors (mutations). Eventually, the cell divides into two daughter cells, in a process called binary fission. The cell size at which division occurs is dependent on the growth conditions \cite{grossman_changes_1982,nelson_penicillin_2000,donachie_cell_1976} (with cells growing on richer nutrients being larger). However, the exact process by which cell division is triggered remains somewhat mysterious, even after half a century of research \cite{donachie_2003,osella,jun,harris,wallden}. Bacteria are able to reproduce at impressive rates: {\em E. coli } can double its population every 20 minutes, under optimal conditions. This means that very large population sizes can be achieved within a few hours in the lab; population densities of $\sim 10^9$ cells per ml of culture medium are usual in lab experiments. 

Protein molecules make up a major component of biomass: typically, $\sim$55\% of the dry mass of a bacterial cell consists of protein \cite{schaechter_microbe._2006}. The bacterial DNA sequence contains several thousand genes ($\sim$ 4000 for {\em E. coli}), each of which encodes a specific protein molecule. Gene expression is the process by which the DNA-encoded instructions for making  a particular protein are  first transcribed into a messenger RNA molecule,  which is  then translated, i.e., used to build a chain of amino acid molecules that folds into a protein molecule. In response to changes in environmental conditions, or signals, a bacterial cell can turn on or off the production of particular protein molecules; this is known as gene regulation. Interestingly, some proteins, known as transcription factors, turn on or off genes that encode other proteins.  This leads to networks of interactions among genes, the properties of which (such as modularity \cite{alon_book}) have attracted significant interest among statistical physicists. Gene regulatory networks are especially interesting because the transcription factors that control them are often present in only a few molecules per cell, leading to stochasticity in the behaviour of the regulatory network (e.g. switching between alternative stable states \cite{gardner,ozbudak}; for theoretical models see, e.g. \cite{warren2005,morelli2008, morelli2009,venegas_2011,visco_prl,visco_pre}).

Many bacterial cells  can also engage in self-propelled motion, which is mediated by various appendages external to the cell. For example, bacteria may swim in liquid media by rotation of whip-like flagella, or crawl on solid surfaces using needle-like appendages called pili \cite{swarm_review,motility_review,berg_book,schaechter_microbe._2006,ob:Brock}.  Bacterial motility has already attracted much interest among physicists; topics of particular focus have included the statistics of suspensions in which bacteria stochastically change direction in response to chemical gradients or local density \cite{cates_review,cates_review2}  and the hydrodynamics of bacterial swimming motion \cite{lauga_review}. Bacterial motion has also inspired a recent surge of work on the  collective behaviour of self-propelled colloidal particles \cite{howse,stenhammar,linek,chucker}.

\subsection{Statistical physics of bacterial growth}
Bacterial growth is  of interest to statistical physicists for several reasons. First, the process of division into daughter cells is a branching process with somewhat stochastic timing; the time between successive bacterial divisions is  a random variable with a rather broad distribution \cite{jun,wallden,kiviet,kennard}. For {\it E. coli}, this causes a loss of synchrony between division events in sister lineages within about 10 generations \cite{hoffman_synchrony_1965,cutler_synchronization_1966}. Stresses such as exposure to some antibiotics or to ultraviolet radiation can interfere with the division process, leading to long, filamentous cells. Even in the absence of stress, bacterial populations can contain small sub-populations of non-growing cells, or "persisters", which tend to be resistant to antibiotic treatment \cite{balaban_2004,balaban_problem_2013}.  

Second, growth of bacteria in close proximity to one another leads to mechanical interactions, which can be thought of as pushing, or excluded volume effects. This is relevant when bacteria grow in dense populations such as colonies on semi-solid surfaces or biofilms on solid surfaces (see Section \ref{sec:hetgrowth}). Mechanical interactions between bacteria  lead to a number of interesting phenomena, including phase separation of cells with different surface properties \cite{PGhosh2015}, segregation of an expanding population into sectors of genetically identical bacteria \cite{hallatschek_2007,hallatschek_2010,ali_reproduction-time_2012}, quasi-nematic ordering \cite{drescher_2016} and competition for space between lineages \cite{lloyd_competition_2015}. Mechanical interactions between bacteria and their environment  can also lead to interesting effects \cite{volfson_biomechanical_2008}, for example a transition from 2d to 3d growth as a bacterial colony grows on a semi-solid agar gel \cite{grant_2014}.

Third, because bacteria reproduce rapidly, they also undergo rapid genetic evolution. The process of evolution involves the random generation of cells with mutations in their DNA, due to mistakes in DNA replication, and their proliferation within the population, starting from initially very small numbers. Bacterial evolution is now widely recognised as an important  testbed for evolutionary theory, since it allows lab experiments to be carried out on short timescales  (typically days-weeks) \cite{west_review,lenski_review}. Understanding how bacterial populations evolve is also a pre-requisite for our ability to mitigate against the emergence of antibiotic-resistant infections \cite{amr_review}.

In the remainder of this review, we highlight in more detail a number of interesting phenomena that are associated with various modes of bacterial growth, and for which statistical physics models have been developed. We divide this discussion into two parts: in section \ref{sec:homgrowth} we consider growth in a homogeneous, well-mixed environment, while in section \ref{sec:hetgrowth} we discuss growth in spatially structured environments.

\section{Bacterial growth in a spatially homogeneous environment}\label{sec:homgrowth}

\subsection{Bacterial growth experiments and population dynamic equations}\label{sec:shm}

In the laboratory, bacteria are often grown in liquid suspension under well-mixed conditions. Here we give a brief overview of the typical experimental techniques involved and the types of equations used to describe the resulting population dynamics.

\subsubsection{Growth in a batch culture}

\begin{figure}
\includegraphics[width=17cm]{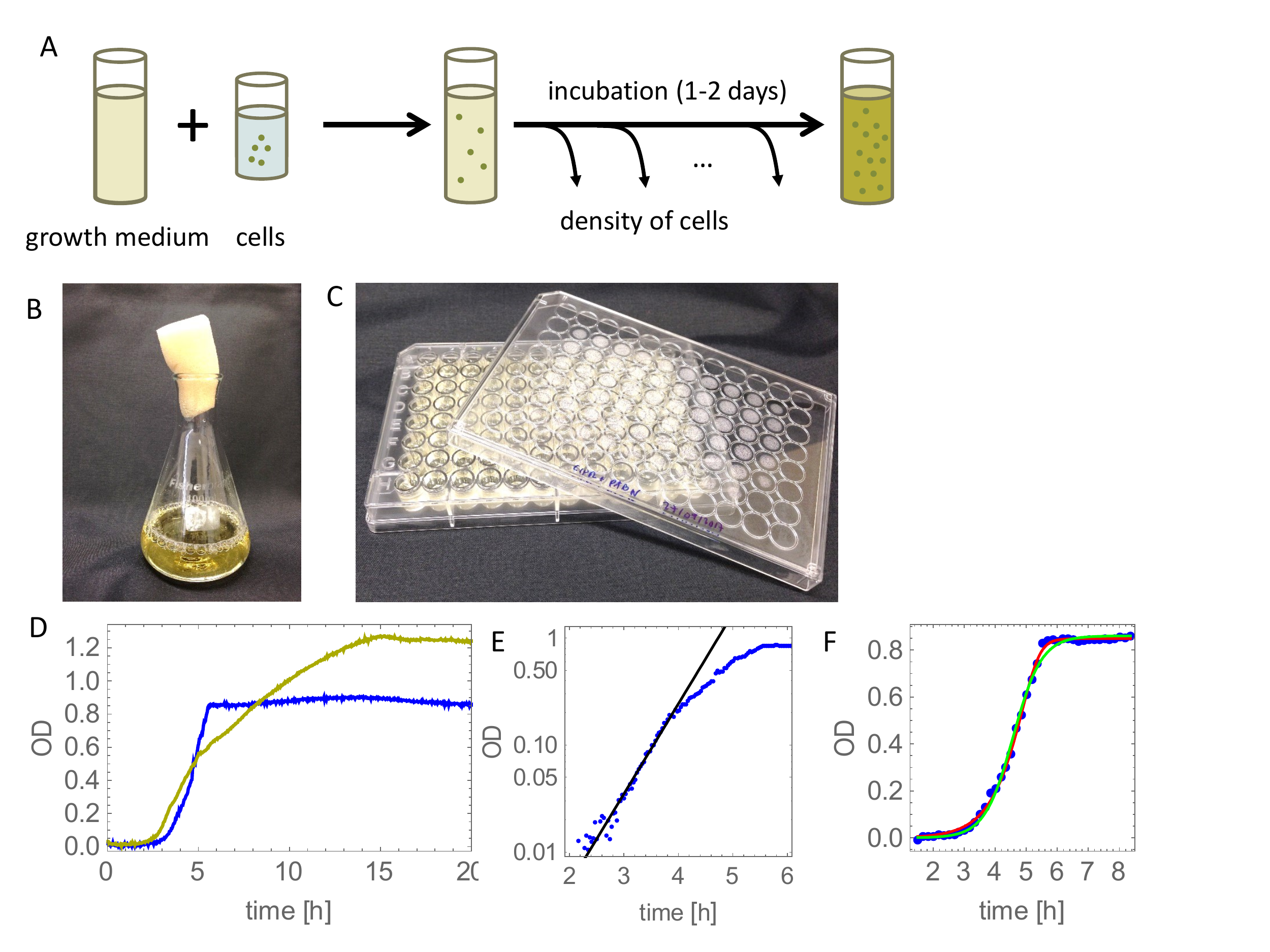}
\caption{\label{fig:growthcurves}(A) Sketch of a typical batch culture bacterial growth experiment. (B-C) Containers used to grow bacteria: a large-volume flask (100ml), and a microtiter plate with 96 individual culture wells of volume $\sim 400\mu$l. (D) Measured growth curves for {\it E. coli} strain MG1655 in simple- (``rich'' MOPS: glucose, aminoacids, nucleotides, salts (blue curve)) and complex-nutrient medium (LB broth, yellow curve). "OD" is a measure of the turbidity of the suspension; see footnote to main text. 
The MOPS medium was created by mixing 100ml M2101, 100ml M2103, 200ml M2104 (Teknova), 10ml 0.132M  K2HPO4, 1g glucose, and double-distilled, autoclaved water to a total volume of 1000ml. The LB medium consists of 25g of LB powder (Fisher): tryptone, yeast extract and NaCl, dissolved in 1000ml of distilled water and autoclaved. 200$\mu$l of the medium was added to each well of a 96-well plate (panel C), inoculated with $1\mu$l of PBS-washed overnight culture of {\it E. coli}, and incubated at 37C in a BMG FLUOstar plate reader for 24h. OD was measured every 2mins with shaking for 20s prior to each measurement.
(E) The exponential growth model (Eq. (\ref{eq:exp_growth}), black curve) fits the experimental MOPS curve from panel D for low bacterial densities. Fitting to the data in the range $t=1.5\dots 3.5$h gives an exponential growth rate $r=1.94$h$^{-1}$. (F) Comparison between different models and an experimental growth curve for growth in rich MOPS: the logistic growth model (Eq. (\ref{eq:exp_growth3})) is shown by the green line and the Monod growth model (Eq. (\ref{eq:exp_growth5})) is shown by the red line. The best-fit maximum growth rate is 2.2h$^{-1}$ (logistic growth) and 2.1h$^{-1}$ (Monod growth). }
\end{figure}

Fig. \ref{fig:growthcurves}A illustrates a typical setup for what is known as a ``batch culture'' growth experiment. A small number of bacteria are inoculated into a well-shaken container  filled with liquid nutrient medium (Fig. \ref{fig:growthcurves}B shows a large-volume flask; Fig. \ref{fig:growthcurves}C shows a 96-well microplate which can be used to perform multiple simultaneous smaller-volume experiments). Over a period of $\sim$1 day, the density of bacteria $n(t)$ is measured (usually by determining the turbidity of the suspension \footnote{For a bacterial suspension, turbidity is usually referred to as ``optical density'', or OD. The OD has been shown to correlate well with the biomass density in the sample  \cite{volkmer_condition-dependent_2011}. Other techniques to measure bacterial density include spreading the suspension on an agar gel of nutrient media, incubating and counting the resulting colonies, or direct counting of cells using a Coulter counter or flow cytometer.}) and the results are plotted as a function of time $t$. Typical results are shown in Fig. \ref{fig:growthcurves}D. These ``growth curves'' have a characteristic shape: an initial period, known as the lag phase, in which no growth is detected, followed by a period of exponential growth (known as the exponential phase), followed by a slowing down and eventual cessation of net growth, known as the stationary phase. It is generally stated that the lag phase happens because the bacteria need to adjust to the liquid medium (having typically been stored under different conditions), while stationary phase happens when the population exhausts its nutrient supply, or builds up  waste products. However, the details of what happens during the lag and stationary phases remains a topic of active research \cite{rolfe2012,kolter}.

Simple equations can be used to describe the results of a batch culture growth experiment. Assuming initially that the nutrients are unlimited, the dynamics of the bacterial population can be modelled as
\bq
	\frac{dN(t)}{dt} = r N(t), \label{eq:exp_growth}
\eq
where $N(t)$ is the number of bacteria at time $t$ and $r$ is the per-bacterium replication rate. Eq. (\ref{eq:exp_growth}) predicts that the population grows exponentially
\bq
	N(t) = e^{r t} N(0) = 2^{t/T} N(0) ,\label{eq:exp_growth2}
\eq
where $T=\ln 2/r$ is the doubling time, defined by $N(t+T)=2N(t)$. For  {\it E. coli} under optimal conditions (rich nutrient broth, 37\degree C), $T\approx 20$ min which gives $r\approx 2.1$ h$^{-1}$. Eq. (\ref{eq:exp_growth}) is appropriate for relatively large populations ($\gg 10^3$ cells). For smaller populations, it may be important to consider that replication is not a continuous process but occurs as discrete events which may be synchronous \footnote{In a population starting from a single bacterium, division events in different cells occur quasi-synchronously for about the first 10 generations \cite{hoffman_synchrony_1965,cutler_synchronization_1966}.}. In some cases, it may be more convenient to use as the dynamical variable the total biomass of the population rather than the number $N$ of bacteria. The total biomass obeys an identical equation to Eq.~(\ref{eq:exp_growth}), but it is a continuous quantity which is unaffected by discrete cell division events \cite{schaechter_growth_1962}.

Equation~(\ref{eq:exp_growth}) provides a good model for the exponential phase of growth of a bacterial population, as we show in Fig. \ref{fig:growthcurves}E. However, it does not capture the transition to the stationary phase, where the population saturates. A simple way to  capture this saturation is to use instead a logistic growth equation \cite{murray_mathematical_2002}
\bq
	\frac{dN}{dt} = r N(1-N/K),  \label{eq:exp_growth3}
\eq
where $K$ is the maximal population size (or carrying capacity) and the  term $(1-N/K)$ decreases the effective growth rate when $N$ becomes large, mimicking the effect of nutrient depletion or toxic waste product buildup. The solution of  Eq.~(\ref{eq:exp_growth3}), $N(t) = N(0) e^{r t}/[1+(N(0)/K)(e^{r t}-1)]$, does indeed saturate, as we show in Fig. \ref{fig:growthcurves}F. This model is in quite good agreement with measured growth curves for experiments in simple nutrient media (Fig. \ref{fig:growthcurves}F) \footnote{The sharp-eyed reader will note that the solution of the logistic equation (\ref{eq:exp_growth3}) is not in perfect agreement with the MOPS growth curve from Fig. \ref{fig:growthcurves}F, and cannot replicate the LB growth curve from Fig. \ref{fig:growthcurves}D. The shapes of growth curves can in general be more complicated than suggested by these simple models, especially where there is more than one growth-limiting nutrient \cite{kk_egli_1998,egli_1991}. More complicated models, such as those that use  density-dependent growth functions  \cite{baranyi_modeling_1993,buchanan_when_1997,pruitt_mathematical_1993,zwietering_modeling_1990}, have been developed to try to achieve better fits.}.

Saturating population growth can also be modelled in a more biologically consistent way by  including the dynamics of the nutrient explicitly in the equations. The classic equation for the nutrient-concentration dependent growth of a bacterial population is:
\bq
	\frac{dN}{dt} =  \left(\frac{r_{\rm max} s}{K_{\rm s}+s}\right) N, \label{eq:exp_growth4}
\eq
where $s$  is the nutrient concentration,  $r_{\rm max}$ is the maximal per-cell growth rate and $K_{\rm s}$ is the nutrient concentration at which the growth rate is half-maximal. In Eq.~(\ref{eq:exp_growth4}), the per-cell growth rate is described by a ``Monod function'' \cite{monod_growth_1949}:
\bq
	g(s)=r_{\rm max} s/(K_{\rm s}+s), \label{eq:Monod}
\eq
which depends linearly on the nutrient concentration $s$ for low nutrient concentrations, but  becomes independent of the nutrient  as $s \to \infty$. This captures the fact that for high nutrient concentration, growth is limited by the bacterium's capacity to import and use the nutrient, rather than by the availability of the nutrient in the environment. Eq.~(\ref{eq:exp_growth4}) must be coupled with a dynamical equation for the nutrient concentration:
\bq
	\frac{ds}{dt} =   -\gamma  \left(\frac{r_{\rm max} s}{K_{\rm s}+s}\right) N,  \label{eq:exp_growth5}
\eq
where $\gamma$ is  a yield coefficient, describing the number of units of nutrient that are consumed to produce one bacterium (divided by the volume). 

Numerical solution of Eqs.~(\ref{eq:exp_growth4}) and (\ref{eq:exp_growth5}) predicts that the bacterial population size saturates as the nutrient runs out. This solution agrees well with experimental data (Fig. \ref{fig:growthcurves}F), although it is typically not significantly better than the solution of the logistic growth equation (\ref{eq:exp_growth3}) \footnote{In addition, the Monod relation (\ref{eq:exp_growth4}) has the rather unsatisfactory feature that it is an {\em ad hoc} function, rather than being derived from any underlying model of the cell's biochemistry. Because of this, attempts have been made over many years to develop more complex nutrient-dependent growth equations, which take into account features such as population-size dependence \cite{contois_kinetics_1959}, temperature \cite{ratkowsky_model_1983}, multiple nutrients \cite{kompala_investigation_1986}, pH \cite{presser_modelling_1997}, and the thermodynamic driving force for the biochemical growth reaction \cite{jin}.}.

\subsubsection{Growth in a chemostat}\label{sec:chemostat}

\begin{figure}
\includegraphics*[clip, trim = 0cm 7cm 0cm 0cm, width=14cm]{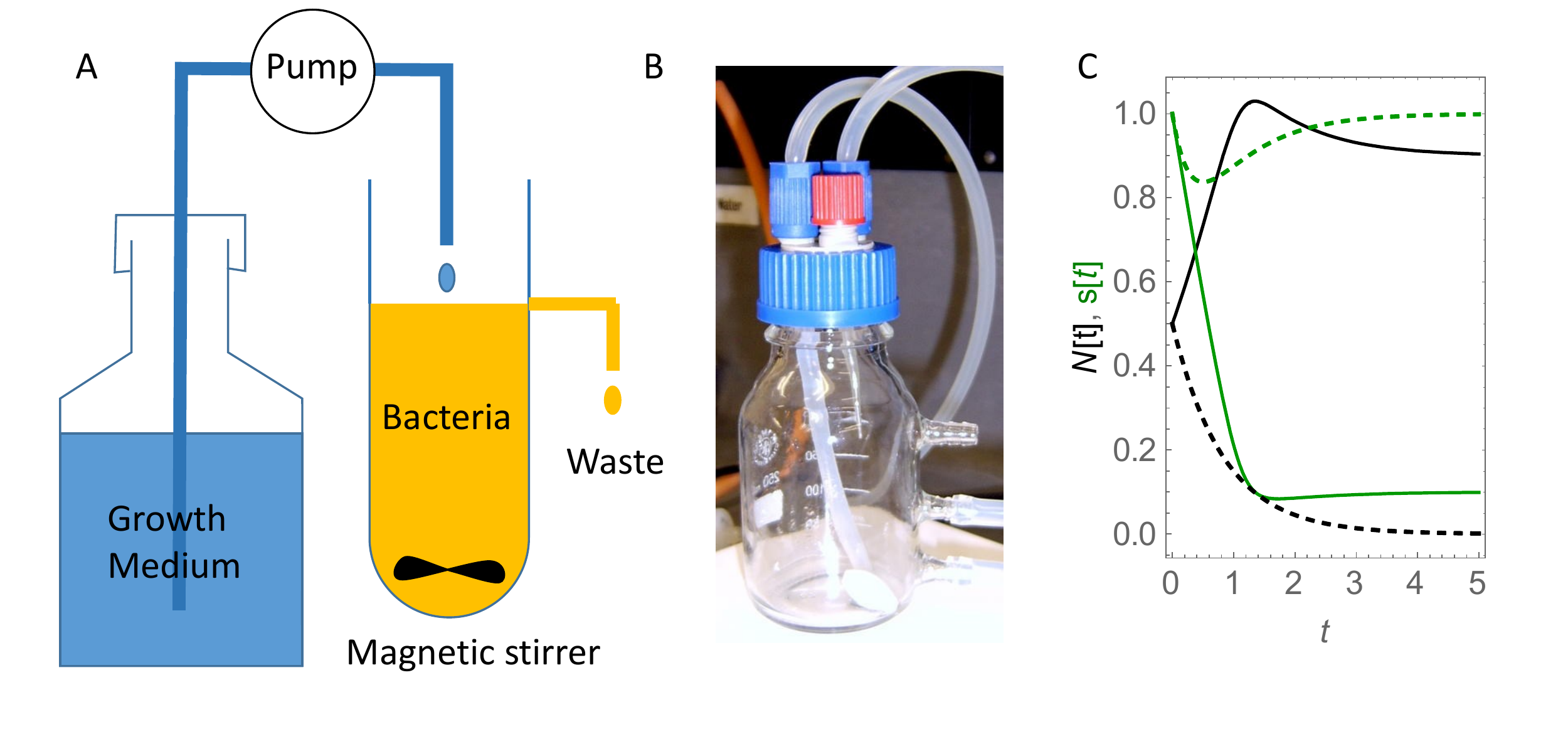}
\caption{\label{fig:chemostat}(A) Schematic illustration of a chemostat. (B) Photograph of a simple chemostat consisting of a glass flask with tubing for media delivery, aeration, and waste removal, and a magnetic bead for stirring the flask contents. (C) Example curves $N(t),s(t)$ (black and green, respectively), obtained by numerical simulation of Eqs. (\ref{eq:chemo1}) and (\ref{eq:chemo}) for $\gamma = 1, K_s = 0.1, s_0 = 1, r_{\rm max} = 2$ and $d=1$ (solid lines) and $d=3$ (dashed lines). The simulations are initiated with $N(0)=0.5$ and $s(0)=1$ (in arbitrary concentration units). $d_{\rm crit}=1.82$ for this set of parameters. For the solid lines, $d<d_{\rm crit}$ and a stable bacterial population is maintained in the chemostat (the solid black line reaches a non-zero steady state); for the dashed lines, $d>d_{\rm crit}$ and the population is ``washed out'' (the dashed black line goes to zero).} 
\end{figure}

The batch culture setup shown in Fig.  \ref{fig:growthcurves} is not the only way to perform a well-mixed bacterial growth experiment. An alternative approach is to use a chemostat:  a well-mixed vessel in which fresh nutrient medium is supplied from a reservoir at a constant flow rate, and the contents of the vessel (bacteria and spent medium) are removed at the same rate, so as to keep the volume constant (Fig. \ref{fig:chemostat}A and B). In the chemostat, one achieves a steady-state population in which the rate of bacterial replication is matched by the rate of removal of bacteria.

The dynamics of bacterial growth in a chemostat can be modelled by making minor modifications to Eqs (\ref{eq:exp_growth4}) and (\ref{eq:exp_growth5}) to account for the inflow of nutrient and the outflow of bacteria plus medium. The resulting equations are \cite{smith_theory_1995}:
\ba
	\frac{dN}{dt} = \left(\frac{r_{\rm max} s}{K_{\rm s}+s}\right)N - Nd,   \label{eq:chemo1}\\
	\frac{ds}{dt} =  -\gamma \left(\frac{r_{\rm max} s}{K_{\rm s}+s}\right)N + s_0d - sd, \label{eq:chemo}
\ea
where $d$ is the rate of fluid flow into and out of the chemostat and $s_0$ is the concentration of nutrient in the reservoir. These equations have the following steady-state solution:
\ba
	N^* &=& \frac{r_{\rm max} s_0 -d (K_{\rm s}+s_0)}{\gamma  (r_{\rm max}-d)}, \\
	s^* &=& \frac{d K_{\rm s}}{r_{\rm max}-d},
\ea
for $d<d_{\rm crit}=(r_{\rm max} s_0)/(K_{\rm s}+s_0)$, and $N^*=0,s^*=0$ if the flow rate is larger than $d_{\rm crit}$. Therefore, growth in the chemostat is possible only if the flow rate is lower than the maximum growth rate of the bacteria\footnote{The same effect of ``washing out'' of the population with a high dilution rate can also be observed in a simpler, logistic-like model without explicit nutrients:
\bq
	\frac{dN}{dt} = r N(1-N/K) - d N,
\eq
which has steady state solution $N^*=K(1-d/r)$ for $d<r$, and $N^*=0$ otherwise.}. Fig. \ref{fig:chemostat}C shows example plots of the bacterial density and the nutrient as a function of time, predicted by Eqs. (\ref{eq:chemo1}) and (\ref{eq:chemo}) for two values of $d$, above and below $d_{\rm crit}$.

The chemostat equations (\ref{eq:chemo1}) and (\ref{eq:chemo}) can be extended to predict the dynamics of multiple competing or cooperating populations (see Section \ref{sec:ex1.1}), populations preyed on by viruses, evolving populations, etc \cite{ferenci,butler,bohannan,khatri}, providing a well-founded mathematical model for a host of ecological scenarios. Many of these models have mathematically interesting solutions (showing, for example, oscillatory dynamics \cite{smith_theory_1995,xia_transient_2005}).

\subsubsection{Growth of small populations}
The models which we have discussed so far are all deterministic; they represent the dynamics of large bacterial populations, for which fluctuations in the population size are negligible. Recently, however, it has become possible to study the dynamics of small bacterial populations using microfluidic devices coupled with microscopy \cite{mothermachine,balaban_2004,keymer}. For example, microfluidic chemostats have been constructed in which the population size is $10^2-10^4$ bacteria \cite{balagadde_long-term_2005,long_microfluidic_2013}. Here, fluctuations in population size become important and stochastic models are needed. The birth-death process provides a natural way to model such a population.  If we assume that bacterial reproduction and death (removal from the system) are Poisson processes with rates $r$ and $d$, then  we can write the following Master equation for the probability $P(N,t)$ that $N$ bacteria are present at time $t$:
\ba
	\frac{dP(N,t)}{dt} = &-&(r+d)NP(N,t) + r(N-1) P(N-1,t) 
    \nonumber \\ &+& d (N+1)P(N+1,t). \label{eq:pstoch}
\ea
The statistical properties of such birth-death processes have been well studied \cite{athreya_branching_2004,nowak_evolutionary_2006,durrett_branching_2015}. One can think of this process as a biased random walk in the space of $N$, the population size, with the strength of the bias being given by $r-d$. If $r<d$ then the removal rate exceeds the birth rate and one expects the population to become extinct within a finite time (i.e. to reach the absorbing state at $N=0$). On the other hand, if $r>d$, then on average the population increases exponentially, $N\sim \exp[(r-d)t]$, but in any given realisation of the dynamics there is a non-zero probability
\bq
	\rho_{N_0} = (r/d)^{N_0}
\eq
that the population will become extinct. This probability decreases exponentially with the initial size $N_0$ of the population since each of the initial cells can go extinct with probability $\rho_1=r/d$ \cite{durrett_branching_2015}. In the critical case where $r=d$ the population fluctuates randomly (as an unbiased random walk in $N$) and will eventually become extinct, but the average time to extinction is infinite. 

Branching and birth-death processes similar to that of Eq. (\ref{eq:pstoch}) have been applied to model bacterial evolution. A classical example is the Luria-Delbr\"{u}ck model \cite{luria_mutations_1943}, or, more precisely, its stochastic version by Lea and Coulson \cite{lea_distribution_1949}. This model predicts the distribution of the small number of mutant bacteria in a large growing population of wild-type (unmutated) bacteria. Comparing the experimentally observed distribution for this quantity with the model prediction is a standard method for estimating mutation probability in bacteria (this is known as a ``fluctuation test'', see \cite{lea_distribution_1949}). For recent developments in this field, see e.g. Ref. \cite{nicholson_universal_2016}.


\begin{center} \noindent\rule{4cm}{0.4pt} \end{center}
\vspace{5mm}

\noindent
In the next two sections, \ref{sec:ex1.1} and \ref{sec:switch}, we review several pieces of recent work in which the models described above are extended to study more complex situations: specifically, noise-driven oscillations in small bacterial populations, and populations of bacteria that switch stochastically between different states.

\subsection{Example 1: Noise-driven oscillations in bacterial populations}\label{sec:ex1.1}
The chemostat, described in Section \ref{sec:chemostat}, is designed to achieve a steady state of growth for a large bacterial population, by supplying fresh medium at the same rate as spent medium (plus bacteria) is removed. In the natural environment, however, bacteria may experience conditions that are very different to those of a chemostat. The population size may be small, as discussed above (e.g. for bacteria inhabiting the spaces between soil granules, or growing inside a human or animal host cell), nutrient supply may be unpredictable, and bacteria may be removed from the system not just by dilution but also by death due to  viral predation or host immune response. Bacteria may also be retained in the system if they adhere to a surface. Extending the chemostat equations (\ref{eq:chemo1}) and (\ref{eq:chemo}) to include these factors reveals interesting predictions, one of which is that noise-driven stochastic oscillations may be a common feature of small bacterial populations in the natural environment \cite{khatri}.

To see this, let us start by analysing the deterministic  chemostat equations (\ref{eq:chemo1}) and (\ref{eq:chemo}), modified to allow for unequal rates of nutrient supply and removal, and for bacterial death. These are:
\ba
	\frac{dn}{dt} =  f(n,s) = \left(g(s)-d\right)n,   \label{eq:bhav1}\\
	\frac{ds}{dt} =   h(n,s) = -\Gamma g(s)n + b - Rs, \label{eq:bhav2}
\ea
where $n = N/V$ is the bacterial density (with $N$ being the number of cells and $V$ the chemostat volume), $s$ is the nutrient concentration, the growth rate $g(s)\equiv r_{\rm max} s / \left(K_{\rm s}+s\right)$, $d$ is the rate of bacterial removal from the system (by death or dilution), $b$ is the rate of nutrient supply, $R$ is the rate of nutrient removal and $\Gamma \equiv V \gamma$ is the yield coefficient. Equations (\ref{eq:bhav1}) and (\ref{eq:bhav2}) have a single (non-trivial) fixed point  at $ n^* = \left(b-RdK_{\rm s}/(r_{\rm max}-d)\right)/(\Gamma d)$ and $s^* = dK_{\rm s}/(r_{\rm max}-d )$. This solution is independent of the volume of the system because the model described by Eqs. (\ref{eq:bhav1}) and (\ref{eq:bhav2}) is deterministic. Linear stability analysis reveals how the system approaches this fixed point \cite{Strogatz}. If the eigenvalues of the Jacobian matrix $\mat{J}$
\begin{equation}\label{Eq:jac}
\mat{J}=\left(
             \begin{array}{cc}
               \partial f/\partial n & \partial f/\partial s \\
               \partial h/\partial n & \partial h/\partial s \\
             \end{array}
           \right) = \left(
             \begin{array}{cc}
               g(s)-d & n (dg/ds) \\
               -\Gamma g(s) & -\Gamma n(dg/ds) -R\\
             \end{array}
           \right),
\end{equation}
evaluated at the fixed point $(n^*,s^*)$, are real and negative, then we expect the system to relax monotonically to its fixed point. In contrast, if the eigenvalues are complex with a negative real part then we expect exponentially decaying damped oscillations as the system approaches the fixed point. The matrix $\mat{J}$ evaluated at $(n^*,s^*)$ is given by
\begin{equation}\label{Eq:jacstar}
\mat{J^*}= d\left(
             \begin{array}{cc}
              0 & \beta/\chi \\
               -\Gamma & -\beta - \chi\\
             \end{array}
           \right),
\end{equation}
where we have defined $\beta \equiv (\Gamma n^*/d)(dg/ds)_{s=s^*} = (\Gamma n^*/d)\times [r_{\rm max}K_{\rm s}/(K_{\rm s}+s^*)^2]$, and $\chi \equiv R/d$. The eigenvalues $\lambda$ of $J^*$ are given by $2\lambda/d = -(\beta+\chi)\pm \sqrt{(\beta+\chi)^2-4\beta}$. If $\chi \ge 1$, {\em i.e.} the nutrient removal rate $R$ is greater than the bacterial removal rate $d$ (e.g. because bacteria adhere to a surface), then $\lambda$ is real and negative for any value of $\beta$, and we expect the system to approach the fixed point monotonically.  However, if $\chi < 1$, {\em i.e.} bacteria are removed faster than nutrient (e.g. due to death), then the eigenvalues are complex with negative real part, for $\beta$ values in a range such that $(\beta+\chi)^2 < 4\beta$. This implies that transient oscillations can happen as the system approaches the fixed point. The frequency $\Omega$ of the oscillations is given by the imaginary part of the eigenvalues $\lambda$: $2 \Omega/d = \sqrt{4\beta - (\beta+\chi)^2}$. Fig. \ref{fig:bhav}A shows that numerical simulations of Eqs. (\ref{eq:bhav1}) and (\ref{eq:bhav2}) indeed predict significant oscillations for a set of parameters corresponding approximately to {\em E. coli} growing on glucose (see figure caption for details) \cite{kk_egli_1998,growth_bacteria_book,Cohen1954}. 

\begin{figure}
\centering
\includegraphics[width=0.8\linewidth]{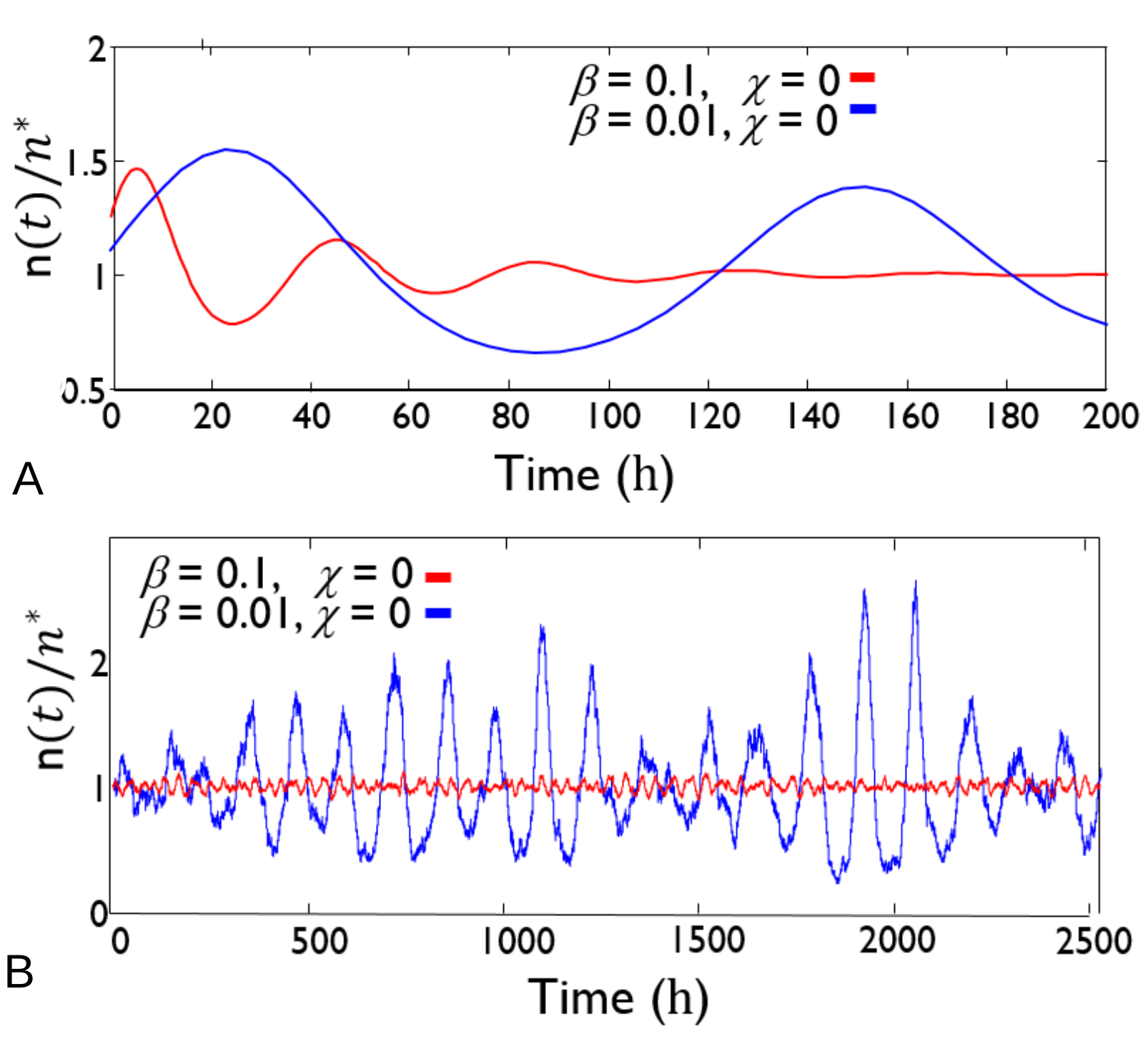}
\caption{Dynamical predictions for the population density of {\em E. coli} bacteria growing on glucose, with parameters $r_{\rm max}=1\mathrm{hr}^{-1}$, $d=0.5\mathrm{hr}^{-1}$, $K_{\rm s}=1\mu\mathrm{M}$ and $\Gamma=1.8\times 10^{10}$ glucose molecules per bacterium . The nutrient inflow rate $b$ is varied, such that we have $\beta=0.1, \chi=0$ (corresponding to $b=0.1\mu$Mh$^{-1}$ and R=0; such that $s^*=1\mu$M and $n^*\approx 10^7$ bacteria per litre) and $\beta=0.01, \chi=0$ (corresponding to $b=0.01\mu$Mh$^{-1}$ and R=0; such that $s^*=1\mu$M and $n^*\approx 10^6$ bacteria per litre). Panel A shows results for the deterministic model, Eqs. (\ref{eq:bhav1}-\ref{eq:bhav2}). Panel B shows results for the stochastic model, Eq. (\ref{eq:lang}), for a system volume of 1ml; i.e. for approximate absolute bacterial numbers of $10^4$ (red) and $10^3$ (blue). Note the different time axes in the two panels. In these simulations, the bacterial densities are much lower than in a typical microbiology lab experiment, and are 1-2 orders of magnitude lower than the bacterial density in seawater, but they are similar to the bacterial density that might be found in drinking water. }
\label{fig:bhav}
\end{figure}

What causes these transient oscillations? Intuitively, they happen because the system builds up  surpluses and deficits of nutrient, relative to the bacterial population density. When there is a surplus of nutrient, the bacterial population grows rapidly and overshoots the amount of available nutrient, leading to a sudden deficit of nutrient, upon which the population decreases rapidly and eventually undershoots the nutrient concentration, leading to a nutrient surplus. A transient nutrient surplus can only happen if excess nutrient  is allowed to accumulate in the system without being washed away; thus the requirement for $\chi < 1$. One can also explain the requirement for an intemediate value of $\beta$, such that $(\beta+\chi)^2 < 4\beta$.  The parameter $\beta$ measures the responsiveness of the bacterial growth rate to changes in the nutrient concentration. For very small values of $\beta$, the growth rate does not respond to changes in nutrient, so transient nutrient surpluses will not translate into bacterial population oscillations. For very large values of $\beta$, the population tracks the nutrient concentration closely, preventing nutrient surpluses or deficits from building up. 

This analysis suggests that, in some situations in the natural environment, bacterial populations whose dynamics is deterministic may undergo transient (damped) oscillations, eventually reaching a non-oscillating steady state. But what happens for very small populations?  It turns out that for small populations stochastic fluctuations due to the birth and death/removal of individual bacteria (demographic noise) drive sustained oscillations in the population density and the nutrient concentration. 

The effects of demographic noise in small bacterial populations can be modelled {\black{in various ways. If the fluctuations due to the noise are expected to be large, then individual birth and death events should be modelled explicitly -- typically these would be modelled as Poisson processes, and simulating using a kinetic Monte Carlo scheme such as the Gillespie algorithm \cite{gillespie1976,gillespie1977}. However, if the fluctuations are expected to be small, they may be approximated by adding stochastic noise terms to the deterministic equations (\ref{eq:bhav1}-\ref{eq:bhav2}). This is the approach taken by Khatri et al \cite{khatri}, leading to }} a set of Langevin equations of the form
\begin{equation}\label{eq:lang}
\frac{d{\vec{x}}}{dt} =  \vec{a} +  \left(\mat{B}\right)^{1/2}\vec{\eta(t)}.
\end{equation}
Here, $\vec{x} \equiv (n,s)$, $\vec{a}\equiv (f(n,s), g(n,s))$ describes the deterministic dynamics, and ${\vec{\eta(t)}}$ is a vector of independent, Gaussian-distributed random numbers with zero mean and variance scaling with the inverse of the system volume (thus, the effects of noise are more important for small system volume). The matrix ${\mat{B}}$ is given by
\begin{equation}\label{Eq:matb}
\mat{B}= \left(
             \begin{array}{cc}
              ng(s)+dn & -\Gamma ng(s) \\
               -\Gamma n g(s) & \Gamma^2 n g(s) + b + Rs\\
             \end{array}
           \right).
\end{equation}
This takes account of the fact that  fluctuations in the bacterial population are coupled to fluctuations in the nutrient concentration, and {\em vice versa}. Equation (\ref{eq:lang}) can be derived via a Kramers-Moyal expansion \cite{vanKampen}; briefly, one expresses the model as a set of  chemical reactions, writes down the corresponding master equation, and Taylor expands it under the assumption that changes in the number of molecules/bacteria due to firing of a single reaction are small \cite{khatri,vanKampen}. {\black{The use of Gaussian noise to model demographic stochasticity is particularly convenient in problems like this one, where the nutrient is represented explicitly. This is because the number of molecules of nutrient is typically far larger than the number of bacteria -- thus the nutrient essentially behaves deterministically. Use of a kinetic Monte Carlo scheme would require one either to simulate nutrient molecules discretely (which would be highly inefficient), or to adjust the kinetic Monte Carlo algorithm to take account of a time-varying, continuous, nutrient concentration (similar to \cite{lu2004}).}}

Fig. \ref{fig:bhav}B shows the results of numerical simulations of Eq. (\ref{eq:lang}), for the same parameter set as in Fig. \ref{fig:bhav}A  (representing {\em E. coli} growing on glucose), but for a system volume of 1ml. These parameters represent a very low bacterial density (similar to that found in drinking water), so that the absolute numbers of bacteria present are $\sim 10^4$ (blue curve) or $\sim 10^3$ (red curve). It is immediately clear that demographic stochasticity has an important effect: the transient oscillations of the deterministic model (Fig. \ref{fig:bhav}A) have been converted into sustained oscillations in the stochastic model. The presence of these oscillations is also clearly visible in the power spectrum  \cite{khatri}. This effect may be widespread for very small bacterial populations; for example, it also happens in a model of a nutrient-cycling bacterial ecosystem with two species which feed on each others' waste products \cite{khatri}. 

These stochastic oscillations are an example of a very  general mechanism that was discovered {\black{by McKane and Newman}} in the context of predator-prey models \cite{McKane2005}, and later found by other statistical physicists in a wide range of models \cite{Lugo2008,Boland2008,Ghose2010,Galla2009,Bladon2010,Alonso2007,McKane2007,Dauxois2009}. In this mechanisms,  the underlying oscillatory modes of a deterministic dynamical system are excited by a source of intrinsic noise (in this case demographic noise), leading to sustained oscillatory dynamics in the stochastic version of the system,  whereas the deterministic system shows only damped oscillations. Thus, this example shows how insights from statistical physics can be important in understanding the behaviour of bacterial populations.

Despite the possible ubiquity of this mechanism, demographic-noise induced oscillations have not yet been observed for bacterial populations. One difficulty is that the effect is strong only if number of bacteria is very small (e.g. Fig. \ref{fig:bhav} shows predictions for $\sim 10^3 - 10^4$ bacteria). The predictions are also for a well-mixed system, while rules out typical experimental methods where small populations are grown as microcolonies on agar plates (see section \ref{sec:hetgrowth}). However, well-mixed conditions for small bacterial populations are starting to be achieved using  in microfluidic chemostats \cite{balagadde_long-term_2005} or microfluidic droplets \cite{jakiela_bacterial_2013}. These techniques should eventually reveal a host of interesting fluctuation-driven dynamical phenomena. 

\subsection{Example 2: Switching bacteria in a switching environment}\label{sec:switch}

\begin{figure}
\centering
\includegraphics[width=0.8\linewidth]{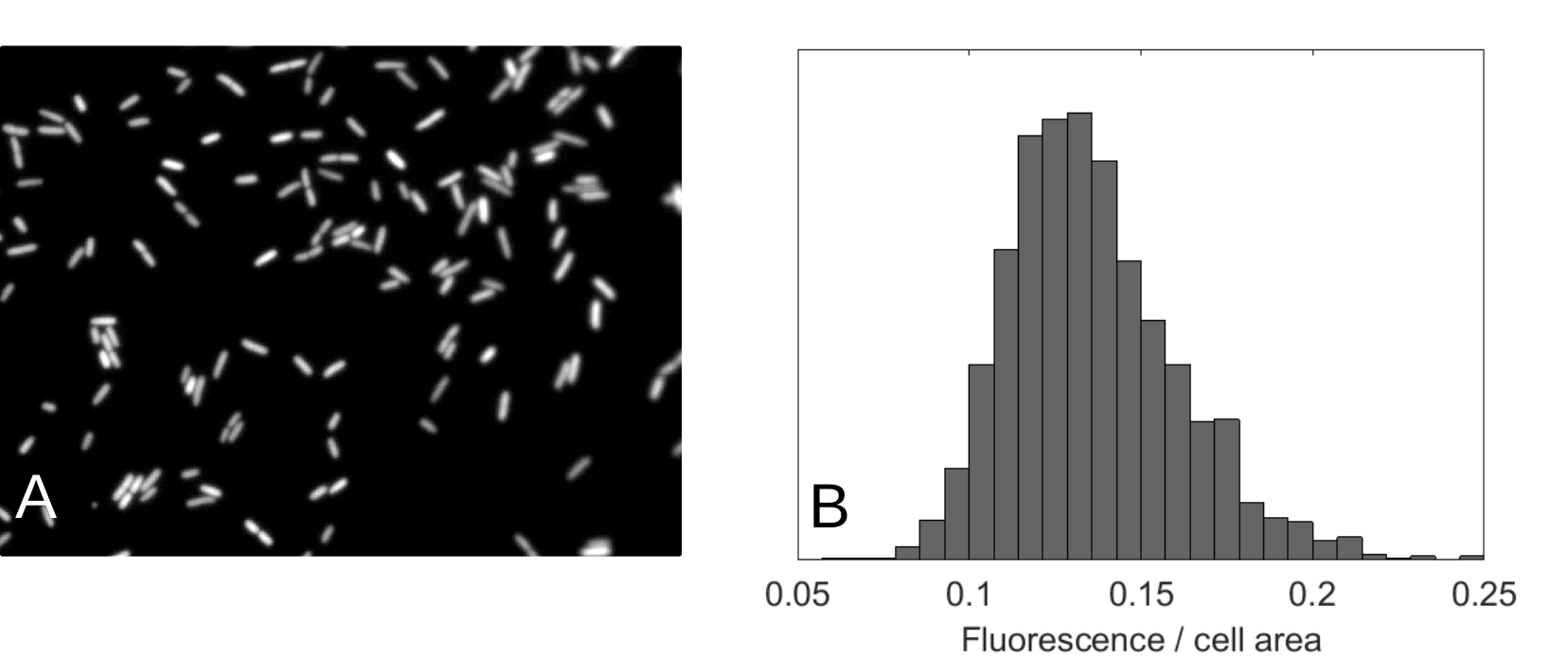}
\caption{{\em E. coli} cells show variability in the expression of a gene encoding a fluorescent protein. A population of  {\em E. coli} cells (strain RJA003, created by P1 transduction from strain MRR \cite{elowitz_2002} into MG1655) was grown in a 1 litre chemostat with dilution rate 0.5h$^{-1}$ on Evans media \cite{evansmedia} supplemented with 50mM glucose. The bacteria expressed cyan fluorescent protein (CFP) from a constitutive (unregulated) promoter (the $P_R$ promoter from phage $\lambda)$. After sampling from the chemostat, cells were kept on ice for $\sim$ 1h prior to being spread on the surface of a 1\% agarose pad and imaged in an epifluorescence microscope. (A) Fluorescence image in the CFP channel showing that individual cells show different levels of fluorescence. (B) Histogram of fluorescence intensities per area, obtained from analysis of many such images (units are arbitrary). The width of the histogram, relative to the mean, provides a measure of the population heterogeneity in gene expression. Data shown courtesy of Joost Teixeira de Mattos, Alex ter Beek, Martijn Bekker and Tanneke den Blaauwen.}
\label{fig:het}
\end{figure}

Up to now, we have mostly assumed that all cells within a bacterial population are identical.  However, in many cases, genetically identical bacteria within a population can show variation in their levels of gene expression (see, e.g., Fig. \ref{fig:het}). In the most striking cases, individual bacteria switch stochastically between very distinct states of gene expression, such that the population contains subpopulations with very different behaviours \cite{woude2004,henderson1999}. The biological function of this stochastic switching is in general not known (and may differ in different cases) \cite{ackermann_review}: suggestions have included evasion of host immune responses \cite{henderson1999,hallet2001,visco}, avoidance of evolutionary fitness valleys \cite{tadrowski}, division of labour among cells in the population \cite{diard} or ``bet hedging'' to ensure survival of the population in an unpredictable environment \cite{arnoldini,seger}. The challenge of explaining the  function of stochastic switching in bacteria has motivated the development of a host of theoretical models, of varying degrees of complexity \cite{lachmann,ishii,thattai,gander,ribeiro,wolf1,kussell,kussell2,visco}. Here, we review perhaps the simplest of these, the case of randomly switching cells in a switching, unresponsive environment \cite{lachmann,ishii,thattai,gander,ribeiro,wolf1}. An elegant statistical physics model for this case was presented by Thattai and van Oudenaarden \cite{thattai}; here we follow their approach, even though it is rather idealistic from a biological point of view.

We suppose that individual bacteria in a population can be in either of two states: a fast-growing state which we label 1 and a slower-growing state, which we label 0.  Bacteria switch stochastically between the two states, with a rate $k_0$ of switching from the fast- to the slow-growing state ($1 \to 0$) and a rate $k_1$ of switching from the slow- to the fast-growing state ($0 \to 1$). This scenario can be described by the following equations for the bacterial population dynamics:
\ba
	\frac{dN_0}{dt} =   - k_{1} N_0 + k_{0} N_1 + \gamma_0 N_0 , \label{eq:sw1}\\
	\frac{dN_1}{dt} =    + k_{1} N_0 - k_{0} N_1 + \gamma_1 N_1 . \label{eq:sw2}
\ea
Here, $N_0$ and $N_1$ denote the numbers of bacteria in each of the two states, and we have assumed that these are large enough to be treated as continuous variables. The first two terms in each equation describe switching between states, 
and the third term describes growth. 

Equations (\ref{eq:sw1}) and (\ref{eq:sw2}) can be reduced to a single, nonlinear dynamical equation by defining the fraction $f$ of the bacterial population which is in the faster-growing state 1 as $f=N_1/N$, where $N=N_0+N_1$. This leads to \cite{thattai}
\ba
	\frac{df}{dt} =   k_{1}  + f[\Delta \gamma -k_1-k_0] - f^2 \Delta \gamma, \label{eq:sw3}
\ea
where $\Delta \gamma = \gamma_1-\gamma_0$ is the difference in growth rate between the two states. The variable $f$ is useful because it provides a measure of the growth rate of the total population, as we can see by summing Eqs. (\ref{eq:sw1}) and (\ref{eq:sw2}):
\ba
	\frac{dN}{dt} =    (\gamma_1+f \Delta \gamma)N . \label{eq:sw4}
\ea
Since $\gamma_1$ and $\Delta \gamma$ are constants, measuring $f$ is equivalent to measuring the total population growth rate. For this reason, $f$ has been referred to as a measure of the ``fitness'' of the population \cite{thattai,visco}. Equation (\ref{eq:sw3}) predicts that the variable $f$ increases in time until it reaches a plateau value which corresponds to the positive root of the quadratic equation $k_{1}  + f[\Delta \gamma -k_1-k_0] - f^2 \Delta \gamma=0$. Thus, although the total population size increases exponentially in time (Eq. (\ref{eq:sw4})), the fraction of the population that is in state 1 approaches a steady state. 

Now let us suppose that the bacterial population lives in a changing environment: specifically, the environment can flip, such that bacteria in the the slow-growing state become fast-growing and {\em vice versa}. Thus, the fraction $f_1$ of bacteria that are in the fast-growing state undergoes a jump: $f_1 \to 1-f_1$. The environment can flip either periodically, or stochastically with a fixed rate \cite{lachmann,ishii,thattai,gander,ribeiro,wolf1}.  The environment is assumed to be ``unresponsive'', in the sense that its behaviour is not coupled to the state of the bacterial population. 

This simple and highly idealized model leads to some interesting results. In particular, one can ask what is the optimal bacterial switching strategy, i.e. the strategy which maximises the total population growth rate. Under what circumstances should bacteria stochastically switch into a slower-growing state, sacrificing fitness in the current environment, in order to be prepared for a change in the environmental state? 

For a periodic environment, it is possible to obtain an analytical solution for the average ``fitness'' $\langle f \rangle$, as a function of the model parameters \cite{thattai}. For a stochastically switching  environment  one has to turn to simulations.  In either case, it turns out that for a certain a range of parameters $(k_0,k_1,\gamma_0,\gamma_1)$ bacterial switching out of the fast-growing state is favourable; thus bacteria can increase their fitness by entering a slow-growing state, in readiness for the next environmental change. Fig. \ref{fig:het2}A shows the results of numerical simulations of Eq. (\ref{eq:sw3}), in an environment that switches at rate 1, either stochastically (as a Poisson process) or periodically. When the rate $k_1$ of switching into the faster growth state is very small, the average fraction $\langle f (k_0)\rangle$ of slowly-growing cells  is peaked at a non-zero value of $k_0$. This implies that it is favourable for cells to switch into the slower growth state at some non-zero rate, whether the environment is stochastic or periodic. However for a larger value of $k_1$, switching into the slower growth state is favourable only in the periodic environment. For even larger values of  $k_1$, switching becomes unfavourable even in the periodic environment \cite{thattai}.

It also turns out that, in regions of parameter space where bacterial switching is favoured, the optimal switching rate matches the switching rate of the environment \cite{thattai}. This prediction has in fact been tested experimentally by Acar {\em et al.}  \cite{acar2}, although using cells of the yeast {\em Saccharomyces cerevisiae} rather than bacteria. In these experiments, yeast cells were engineered to switch stochastically between two states, in which expression of an enzyme for metabolising the nutrient uracil was either on or off. Importantly, the rate of switching could be controlled by addition of a chemical inducer. The yeast cells were grown in a turbidostat (a setup similar to a chemostat, but where nutrient is supplied when the culture reaches a predefined cell density rather than continuously), in which the environment either contained uracil (such that ON cells were fitter than OFF cells) or a toxic analogue of uracil (such that OFF cells were fitter than ON cells). Fig.  \ref{fig:het2}B illustrates this setup: the environment was maintained in state $E_1$ for time $T_1$ before being switched to state $E_2$ for time $T_2$. Yeast cells switched between the two phenotypic states at (controllable) rates $r_{\rm ON}$ and $r_{\rm OFF}$ and proliferated at rates (here labelled $\gamma$) that depended on both the phenotypic state and the environment. Fig. \ref{fig:het2}C and D show results of experiments for a "fast" environment, in which $T_1$ and $T_2$ are relatively short (Fig. \ref{fig:het2}C) and a "slow" environment, in which $T_1$ and $T_2$ are long (Fig. \ref{fig:het2}D). In the fast  environment, rapidly switching cells, with high rates $r_{\rm ON}$ and $r_{\rm OFF}$ (red data points) have, on average, a faster growth rate than slow-switching cells, with lower rates $r_{\rm ON}$ and $r_{\rm OFF}$ (blue data points). The situation is reversed, however, for the slow  environment.

\begin{figure}
\centering
\includegraphics[width=0.8\linewidth]{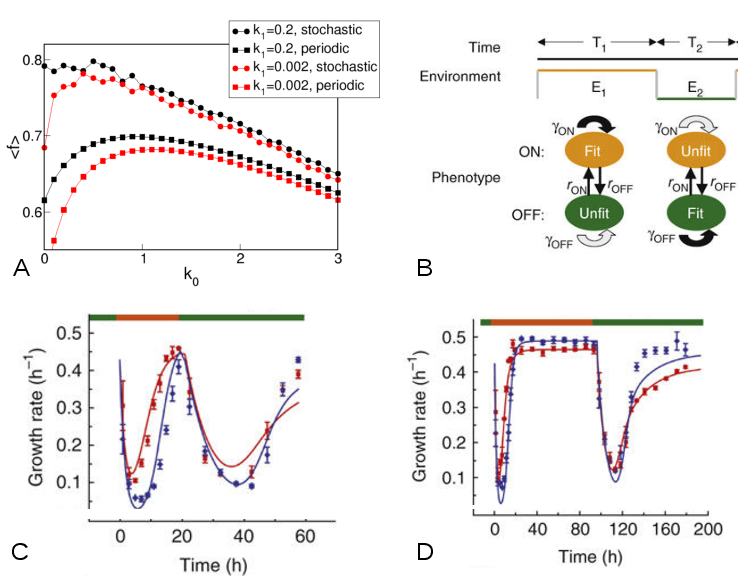}
\caption{(A) Predicted average value of $f$ as a function of the rate $k_0$ of switching into the less favourable state 0. These results were obtained by numerical solution of Eq. (\ref{eq:sw3}), for an environment that switches stochastically as a Poisson process (with average rate 1), or periodically (once per time unit). When the environment switches, we set $f \to 1-f$ in the simulation, keeping all other parameters fixed. When the rate $k_1$ of switching into the faster growth state is very small, the function $\langle f (k_0)\rangle$ is peaked at a non-zero value of $k_0$, implying that it is favourable for cells to switch into the slower growth state at some non-zero rate, whether the environment is stochastic or periodic. However for the larger value of $k_1$ simulated here, switching into the slower growth state is favourable only in the periodic environment. For even larger values of  $k_1$, switching becomes unfavourable even in the periodic environment  \cite{thattai}. (B-D) Experiments with a stochastically switching strain of yeast cells, performed by Acar {\em et al.} \cite{acar2} (images reproduced from Ref. \cite{acar2}). (B) Schematic illustration of the experimental setup. Yeast cells can be in either of two states, labelled ON and OFF. Cells randomly switch between the states at rates  $r_{\rm ON}$ and $r_{\rm OFF}$ which can be tuned by the experimenter. The environment is maintained in state $E_1$ for time $T_1$ before being switched to state $E_2$, which is maintained for time $T_2$. The proliferation rates $\gamma$ of the two cell types depend on the environment; the ON cell type proliferates faster in $E_1$ while the OFF cell type proliferates faster in $E_2$. (C-D) Growth rates measured as a function of time during such an experiment, for cells that switch fast (red; $r_{\rm ON} \sim 0.047$h$^{-1}$, $r_{\rm OFF} \sim 0.035$h$^{-1}$) or slow (blue; $r_{\rm ON} \sim 0.004$h$^{-1}$, $r_{\rm OFF} \sim 0.007$h$^{-1}$). Panel C shows results for an experiment with $T_1=20$h, $T_2=37h$; here the fast-switching cells have on average a higher growth rate. Panel D shows results for $T_1=96$h, $T_2=96h$; here the slow-switching cells have a higher growth rate on average. \label{fig:het2}}
\end{figure}

This example shows that even the relatively simple case in which the switching behaviour of the cells and of the environment is uncoupled can produce non-trivial results, which go some way to explaining the possible advantages of stochastic switching. Real infections or environmental scenarios are of course more complex, and other models have been developed that reflect different aspects of this complexity. For example, statistical physicists have considered the case of an environmental switch which is triggered when the state of the population reaches a threshold (mimicking an immune response) \cite{visco}. With some approximations, this case can be treated analytically and reveals a new possible role for stochastic switching, in which the population composition is modulated so as to avoid triggering the environmental response. In other work, looking at the topic from a different perspective, statistical physics models have been used to investigate the relative benefits of ``blind'' stochastic switching compared to ``responsive'' switching, in which cells detect the state  of the environment and respond accordingly \cite{kussell}. A scenario in which stochastic phenotype switching is advantageous even in a fixed environment has also been considered, in Ref. \cite{tadrowski}.

\section{Spatially structured bacterial populations\label{sec:hetgrowth}}

\begin{figure}
\centering
\includegraphics[width=0.8\linewidth]{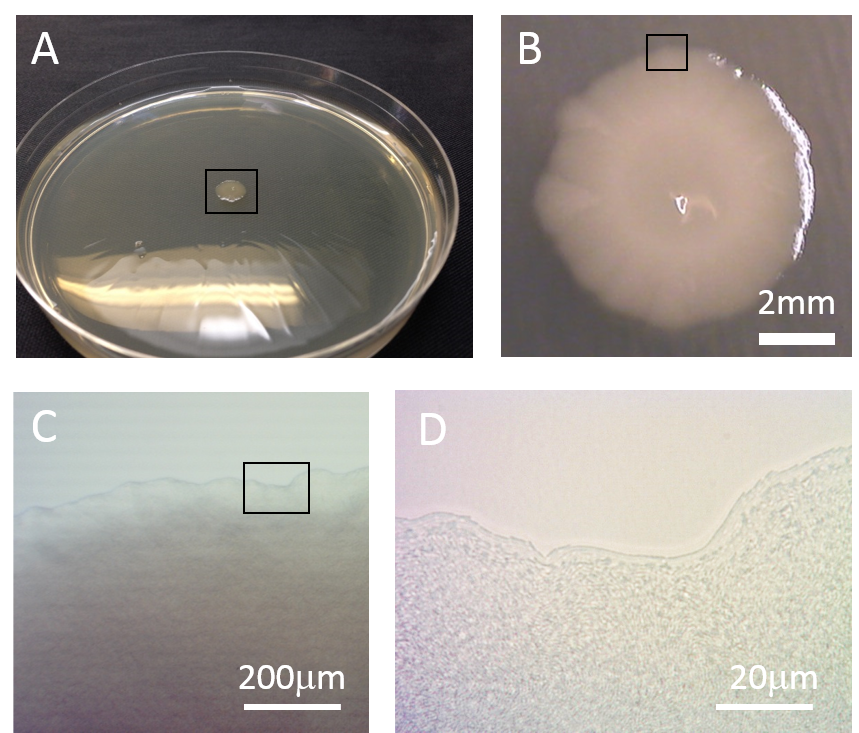}
\caption{(A) A colony of {\it E. coli} growing on the surface of an agarose gel in a Petri dish. The colony is about 7mm wide and less than 1mm thick. (B-D) Successive close-ups of a fragment of the colony's rough border. In (D), individual bacteria may be seen.}
\label{fig:colonies_E_coli_zoom}
\end{figure}

\begin{figure}
\centering
\includegraphics[clip, trim = 0cm 25cm 0cm 0cm, width=0.9\linewidth]{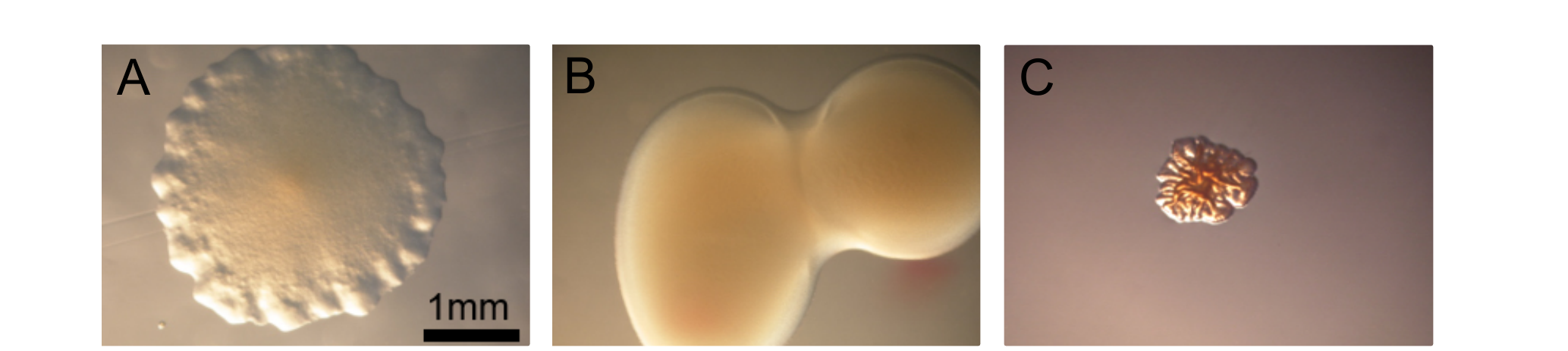}
\caption{(A-C) Colonies formed by genetic variants of {\em Pseudomonas aeruginosa} strain PA01 on the surface of a nutrient agar pad. Images courtesy of Yasuhiko Irie, University of Dayton. (A) A colony formed by the standard (non-mutated) version of this strain. (B): A ``mucoid'' colony formed by a strain that overproduces extracellular polymeric substances. (C) A ``rugose'' colony formed by a ``rugose small colony variant'' (RSCV) strain. RSCV strains often show increased levels of the intracellular signalling molecule cyclic di-GMP. }
\label{fig:colonies}
\end{figure}

In nature, bacterial populations rarely exist as well-stirred, homogeneous, liquid cultures. Instead, imperfect mixing, combined with spatial heterogeneity of the environment (e.g. gradients of food, oxygen, temperature), leads to the emergence of populations which are spatially structured, both genetically and phenotypically \cite{stewart_diffusion_2003, baquero_challenges:_1997}. These structured populations often take the form of dense conglomerates in which bacterial cells interact mechanically with each other. An example is a bacterial biofilm, which is a dense mat of cells attached to a surface \cite{costerton_review,kolter_review}. This might be a solid surface such as a rock or soil particle, or a semi-solid matrix such as food, animal or plant tissue. Biofilms are a source of concern in both medicine and industry because they can cause chronic infections when they form on medical implants, and biofouling when they form on industrial devices \cite{parsek_review, biofouling_review}.


In the microbiological lab, dense, spatially structured bacterial populations are often encountered in the form of "colonies". These colonies arise when individual bacterial cells are dispersed across the surface of a layer of nutrient-containing semi-solid agar gel and allowed to proliferate in an incubator for a day or so (Fig. \ref{fig:colonies_E_coli_zoom}A)\footnote{Agarose is a polymer of agarobiose monomers, whereas agar is a natural product produced by algae that contains a mixture of agarose and agaropectin. Agar is cheaper and is used in plating experiments; agarose is more expensive but is often preferred for microscopy.}. The colonies are visible by eye as small spots of size $\sim 0.5-5$mm on the agar surface; each one contains $\sim 10^8$ or more bacteria, all of which are progeny of a single founder cell (Fig. \ref{fig:colonies_E_coli_zoom}B-D). While Fig. \ref{fig:colonies_E_coli_zoom} shows colonies formed by {\em E. coli}, Fig. \ref{fig:colonies} shows those formed by several genetic variants of the bacterium {\em Pseudomonas aeruginosa} (PAO1 strain). These variants show strong differences in colony appearance due to differences in their production of extracellular polymers which affect cell-cell and cell-surface interactions. More generally, the shape and size of a bacterial colony depend on factors such as the nutrient concentration and the agar gel stiffness as well as the bacterial strain that is used \cite{benjacob_review,benjacob_review2}. 

From a physicist's point of view, the growth of biofilms and bacterial colonies are beautiful examples of self-assembly processes, in which the structural properties of the population are closely coupled with local gradients of nutrient (or, potentially, of signalling molecules or toxic substances such as antibiotics) \cite{amr_review, dockery_2001, nadell,melaugh_2016}.  As with many other statistical physics models, the inclusion of space in models for bacterial population growth leads to many interesting new phenomena. 

Biofilm and colony self-assembly are particularly interesting from a statistical physics perspective because of their connection with the well-established field of interface growth models \cite{bonachela_universality_2011}. Interface growth models fall into a small number of universality classes, with well-defined scaling exponents for the interface roughness as a function of time and system size \cite{krug_origins_1997}. Frustratingly, though, there are few experimental systems \black{\cite{maunuksela_kinetic_1997,takeuchi_universal_2010,takeuchi_growing_2011,takeuchi_evidence_2012}} for which these theoretically-predicted scaling exponents can be measured. \black{The edge of an expanding bacterial colony or the surface of a growing biofilm could provide an excellent system to measure such exponents and may stimulate research into models that involve non-local interactions between remote regions of the interface. Such long-range} interactions can happen in bacterial populations due to the interplay between growth and nutrient/waste diffusion, and the dynamics behind the front caused by physical interactions between growing cells \cite{grant_2014, farrell_mechanically_2013}.

\subsection{Modelling spatially structured bacterial populations}

Many different approaches can be used to model spatially structured bacterial populations, depending on the system being studied and the desired level of physical and biological realism.

\subsubsection{Connected habitats and Fisher-KPP waves}

\begin{figure}
\centering
\includegraphics[clip, trim = 0cm 1cm 4cm 6cm, width=0.9\linewidth]{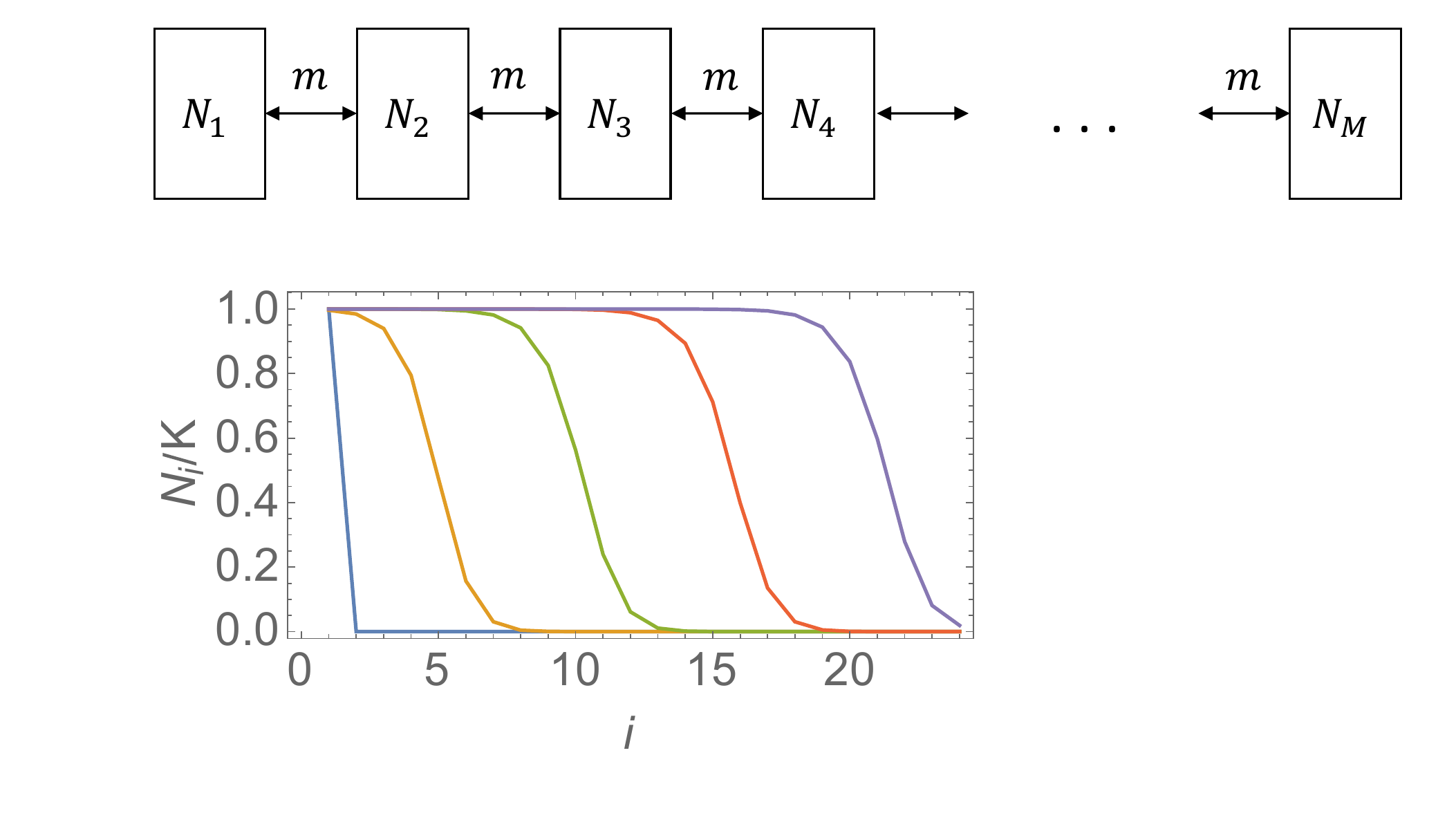}
\caption{\label{fig:fisher}Top: Schematic illustration of a model in which a spatially structured bacterial population is represented as a chain of well-mixed sub-populations connected by migration. In this model, bacterial growth in each compartment is assumed to follow the logistic model (Eq. (\ref{eq:exp_growth3})), and the migration rate per bacterium between neighbouring compartments is assumed to be a constant, $m$. Bottom: The model predicts the emergence of travelling waves of bacteria. The plot shows the results of numerical simulations of Eq. (\ref{eq:compart}), for $M=24,r=1,m=0.1$ and any $K>0$, in which bacteria are initially present only in compartment $i=1$. The curves correspond to times $t=0,7.5,15,22.5,30$ (from left to right).}
\end{figure}

Perhaps the simplest approach is to construct a model that consists of connected well-mixed compartments, between which bacteria can migrate (mimicking motility, diffusion or flow). The dynamics of the bacterial population in each compartment can be described using the same type of equations as in section \ref{sec:shm}, with the additional of coupling terms to describe migration of bacteria between compartments. This kind of approach is appropriate for situations where the local environment of the bacterial population is liquid-like, and the lengthscale over which the environment varies is large. It is often used, for example,  in large-scale models of ocean plankton dynamics \cite{plankton_dynamics}, and in macro-scale models for the treatment of infections with antibiotics \cite{pk_book}.

As a specific example, let us consider a population of bacteria growing in a  chain of $M$ connected compartments (Fig. \ref{fig:fisher}, top). We assume that the bacterial growth dynamics with each compartment can be described by the logistic model
(Eq. (\ref{eq:exp_growth3})), and that the migration rate per bacterium between neighbouring compartments is constant. This leads to the following set of differential equations:
\ba \label{eq:compart}
	dN_i/dt = rN_i(1-N_i/K) + m(N_{i+1} + N_{i-1} -2N_i), \quad \mbox{for}\; 1<i<M  \\
	dN_1/dt = rN_1(1-N_1/K) + m(N_{2} -N_1), \nonumber \\
	dN_M/dt = rN_M(1-N_M/K) + m(N_{M-1} - N_M), \nonumber
\ea 
where $N_i$ denotes the number of bacteria in compartment $i=1,\dots,M$, $r$ is the replication rate, $K$ is the carrying capacity  and $m$ is the migration rate. Let us suppose that bacteria are initially present only in compartment $i=1$. In this case, the population spreads in a wave-like manner to the other compartments, as we show by numerical simulation in Fig. \ref{fig:fisher} (bottom). In the limit of many compartments, assuming a small distance $\Delta x$ between compartments, and setting $m/(\Delta x)^2\to D$, we can rewrite Eq. (\ref{eq:compart}) as a partial differential equation:
\bq
	\frac{\partial n}{\partial t} = D\frac{\partial^2 n}{\partial x^2} + r n\left(1-\frac{n}{K'}\right),
    \label{eq:fisher}
\eq
where $x$ denotes spatial position, $n(x,t)$ is the local density of bacteria and $K'=K/V$ (with units of bacterial density; here $V$ is the volume of one compartment). This is an example of a Fisher-Kolmogorov-Petrovsky-Piscounov (FKPP) equation \cite{murray_mathematical_2002}; its solutions are travelling waves similar to those observed for the discrete case (Fig. \ref{fig:fisher}). Stochastic versions of the  FKKP equation have also been studied \cite{hallatschek_fisher,conlon}, and these may provide a good description of bacterial population dynamics in some circumstances  \cite{hallatschek_pnas}. Related approaches have also been used to model more complex situations, including the spatial expansion of several interacting bacterial populations \cite{venegas-ortiz_speed_2014} and the evolution of resistant bacteria in a drug gradient \cite{hermsen_rapidity_2012,greulich_mutational_2012}.

\subsubsection{Continuum models for dense populations}

\begin{figure}
\centering
\includegraphics[width=1.0\linewidth]{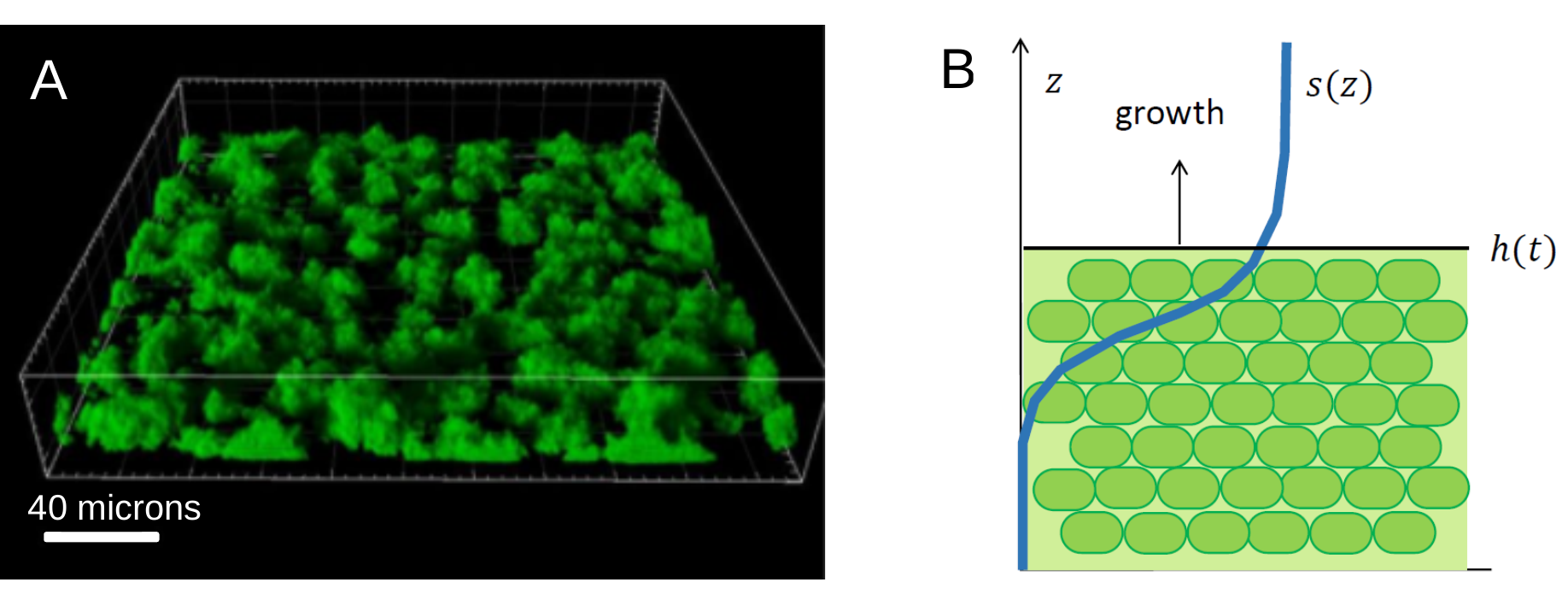}
\caption{\label{fig:1dcolony} (A) Confocal laser scanning microscope image of a biofilm formed by {\em P. aeruginosa} PAO1 grown for 24 hours in a flow cell. Image reproduced from Ref. \cite{kragh_2016}. (B) Illustration of a simple model of a growing biofilm with a flat boundary. The position of the boundary is given by $z=h(t)$ and the nutrient profile is $s(z)$.}
\end{figure}

The situation is different when bacteria grow in a densely packed assembly such as a colony or a biofilm, in which individual cells do not migrate freely. Here, physical interactions between bacteria are likely to be important, and there are also likely to be steep local gradients of nutrient or other chemicals, making it necessary to model chemical concentration fields explicitly. In such cases, one can still use a continuum approach, in which both chemical concentrations and the bacterial population density are represented as continuous fields (as in the FKPP equation), but the equations must be formulated differently \cite{klapper_mathematical_2010,farrell_mechanically_2013}.

As an example, let us consider the growth of a biofilm on a solid surface, as shown in Fig. \ref{fig:1dcolony}A. If we suppose that the bacterial cells are densely packed then it is reasonable to assume the bacterial density $n$ is constant within the biofilm. If we also assume the biofilm is flat (even though this is clearly not a good assumption for the biofilm of Fig. \ref{fig:1dcolony}A!), then the problem becomes 1-dimensional and the bacterial density as a function of the vertical coordinate is a step function of height $h(t)$ (Fig. \ref{fig:1dcolony}B). As the biofilm grows, $h(t)$ increases to accommodate the increase in biomass. The dynamics of $h(t)$ can be written as
\begin{equation}
\frac{\partial h(t)}{\partial t} =  \int_{0}^{h(t)} n g(s(z,t)) dz  \label{eq:bf2}
\end{equation}
where $g(s)$ is the growth rate, which depends on the local nutrient concentration $s$, for example {\em via} a Monod function (Eq. (\ref{eq:Monod})). The dynamics of the nutrient concentration $s(z,t)$ is governed by diffusion into the biofilm and consumption by the bacteria:
\begin{equation}
\frac{\partial s(z,t)}{\partial t} = D\frac{\partial^2 s(z,t)}{\partial z^2}-\Gamma n g(s(z,t))  \Theta(h(t)-z). \label{eq:bf1}
\end{equation}
Here, $D$ represents the diffusion constant of the nutrient (assumed to be the same inside and outside the biofilm, for simplicity), $\Gamma$ is a yield coefficient (nutrient consumed per unit of biomass created) and $\Theta$ is the Heaviside step function.  Depending on the boundary conditions for the nutrient field $s(z,t)$, the choice of growth function $g(s)$ and the parameters $D$ and $\Gamma$, this model can predict linear growth: $h(t)\propto t$, growth that slows down in time: $h(t)\sim \sqrt{t}$, or exponential growth: $\ln [h(t)]\sim t$ \cite{dockery_2001,klapper_finger_2002,farrell_mechanically_2013}.

In the above example, we have assumed a flat biofilm, which allows us to reduce the problem to one dimension. In reality, however, most biofilms have rough surfaces when grown in the typical laboratory flow cell setup (e.g. Fig.\ref{fig:1dcolony}A) \cite{tolker}; some even have  ``mushroom''-like protrusions \cite{klausen}. More realistic continuum models take surface roughness into account by representing the biofilm in two or three dimensions, and also account for spatially varying bacterial density and local pressure within the biofilm  \cite{dockery_2001,klapper_finger_2002,giverso_emerging_2015}. Such models can show interesting phenomena including a ``fingering instability'' \cite{dockery_2001,klapper_finger_2002,wang_shape_2017}, as we discuss in more detail in section \ref{sec:phasesofgrowth}. \black{``Active nematics'' models that take into account orientation of non-spherical cells inside the biofilm can also explain features of bacterial colonies such as the existence of nematic-like defects and micro-domains of locally-aligned cells \cite{doostmohammadi_defect-mediated_2016,you_geometry_2017}
}.

\subsubsection{Individual-based models, on and off-lattice\label{sec:ibm}}

In some situations, it is not appropriate to treat a spatially structured bacterial population as a continuous field; one requires instead detailed spatial resolution at the level of individual cells. This is the case, for example, if one is interested in small populations, heterogeneous populations (e.g. stochastically switching cells, as in section \ref{sec:switch}), or population-level processes that are triggered by single-cell events (as we shall see in section  \ref{sec:buckling}). Models in which the position and state of each bacterial cell is tracked in time are known as individual-based models or agent-based models.

\begin{figure}
\centering
\includegraphics[clip, trim = 0cm 10cm 0cm 0cm,width=0.7\linewidth]{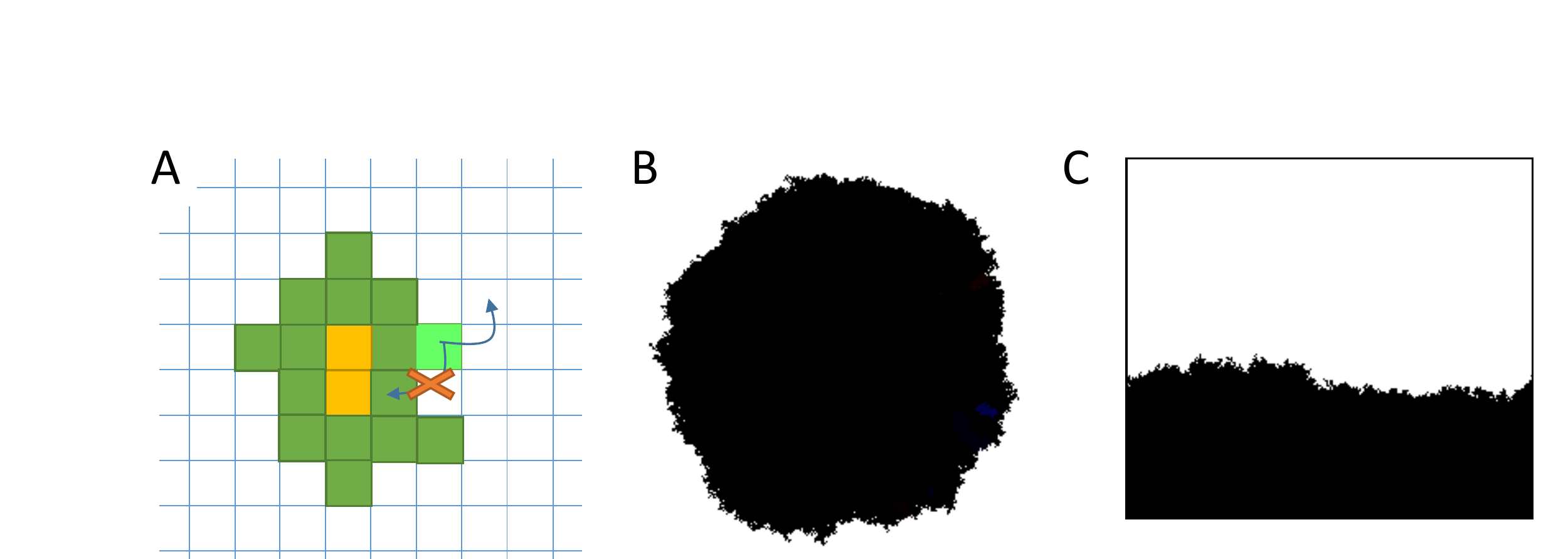}
\caption{\label{fig:2dcolonies}(A) The Eden model on a two-dimensional lattice. Cells that are completely surrounded and cannot replicate are shown in yellow. The bright green cell illustrates the replication rules: it can replicate to a neighbouring empty lattice site (including the diagonal ones in this variant of the model) but not to an occupied one. (B) A snapshot of a simulation of the two-dimensional Eden model in which a cluster of cells ($N=65536$) has grown from a single initial cell. (C) Simulation snapshot for a two-dimensional Eden model in a domain of width $L=250$ that is semi-infinite in the vertical dimension, with periodic boundaries in the horizontal dimension.}
\end{figure}

The simplest form of an individual-based model of a bacterial population is a lattice-based one, in which bacteria occupy sites on a lattice and reproduce into neighbouring lattice sites according to certain rules. A classic example is the Eden model \cite{eden} (Fig. \ref{fig:2dcolonies}A). In the Eden model, lattice sites are either empty or occupied, and an occupied site, or ``cell'', can reproduce if empty sites are available in the neighbourhood (different variants of the Eden model make different assumptions about how the replication rate depends on the number of empty neighbours \cite{plischke_dynamic_1985, kertesz_noise_1988, alves_eden_2012}). Starting from a single occupied lattice site, the Eden model produces a cluster of occupied sites (Fig. \ref{fig:2dcolonies}B) whose interfacial properties fall into the Kardar-Parisi-Zhang (KPZ) universality class \cite{kpz_review}. A particularly important descriptor of an interface is its roughness, defined as the standard deviation of its height fluctuations. If we consider for simplicity a system with the geometry shown in  Fig. \ref{fig:2dcolonies}C (an infinitely long slab of width $L$ sites), then the interface roughness $W$ is given by 
\bq
	W = \sqrt{\frac{\sum_i (h_i - \langle h \rangle)^2}{L}}, \label{eq:roughness}
\eq
where $h_i$ is the vertical height of the cluster of cells at horizontal position $i$. The roughness scales as $W\sim t^\beta$ for short times and $W\sim L^\alpha$ for long times, with the two critical exponents being $\alpha=1/2$ and $\beta=1/3$ for the two-dimensional Eden model. 
The same critical exponents are obtained in an off-lattice version of the Eden model \cite{ali_reproduction-time_2012}.

How well does the Eden model capture the behaviour of real bacterial populations? To our knowledge, interfacial growth exponents have not been measured for flow-cell biofilms like that of Fig. \ref{fig:1dcolony}A. For bacteria growing as colonies on agar gel surfaces, however, surface roughness $W$ has been measured under conditions where the bacteria are non-motile \cite{vicsek1990, wakita}. The results are somewhat mixed: some experiments have produced exponents $\alpha$ and $\beta$ that are different from those of the KPZ universality class (with various suggested explanations \cite{vicsek1990, wakita, bonachela_universality_2011}), while other  experiments have produced exponents consistent with KPZ \cite{gralka2016}. It seems that the jury is still out on whether at least some bacterial colonies fall into the KPZ universality class. For colonies of motile bacteria, or under conditions of low nutrient concentration, more  complicated, fractal-like colony structures can arise \cite{benjacob_review, ohgiwara, fugikawa, ruzicka} and the interfacial growth exponents $\alpha$ and $\beta$ are very different to that of the KPZ universality class (and correspondingly the Eden model).

The most serious limitation of the Eden model and similar lattice models is that growth is restricted to cells that are at the boundary of the cluster, and hence the centre of the cluster is static. This is not a good representation of most bacterial colonies. While bacteria in the centre of a colony do become starved due to insufficient nutrient penetration as the colony becomes large, growth typically occurs in the outer parts of the colony in a layer of considerable thickness (tens of cellular diameters), and within this growing layer elongation and proliferation of bacteria behind the colony edge lead to pushing forces on bacteria that are closer to the edge. A similar picture holds for bacterial biofilms. One can devise lattice models that are somewhat more realistic, by allowing cells in the centre to replicate and push away surrounding cells \cite{waclaw_spatial_2015}. However, off-lattice individual-based models  offer a much greater level of realism. Such models can account for physical interactions between neighbouring bacterial cells and between bacteria and their environment, as well as the dynamics of nutrients, intra-cellular chemical signals, toxins, etc.

In off-lattice individual-based models, individual bacteria (usually modelled as disks in 2D, or spheres in 3D, although rod-shaped cells can also be modelled) move in continuous space and interact via physical mechanisms (e.g. elastic repulsion, friction etc). These simulations are somewhat analogous to molecular dynamics or Brownian dynamics simulations in condensed matter physics. For example, the dynamics of two-dimensional rod-shaped bacteria whose motion is opposed by viscous-like friction can be described by the following equations \cite{farrell_mechanically_2013, farrell_mechanical_2017}
\ba
	d\vec{r}_i/dt = \vec{F}_i/(\zeta l_i),	\label{eq:dr} \\
	d\phi_i/dt = 12\tau_i/(\zeta l_i^3)  \label{eq:dphi},
\ea
where $l_i$ is the length of bacterium $i$, $\vec{r}_i$ is the position of its centre of mass, $\phi_i$ is its angular orientation, $\vec{F}_i$ and $\tau_i$ are the total force and torque acting on it, and $\zeta$ is the friction (damping) coefficient. \black{The dependence on $l_i$ comes out from the assumption that every infinitesimally thin section of the rod experiences a friction force proportional to the local velocity.}
The model must also account for bacterial growth (here, increase in $l_i$ with time) and division. The rate of growth typically depends on  a local nutrient concentration field which is represented on a grid and is updated at each timestep according to a reaction-diffusion equation accounting for diffusive transport and bacterial consumption:
\bq
		\frac{\partial s}{\partial t} = D \nabla^2 s - \gamma \sum_i g(s(\vec{r_i})).
\eq
A variety of models of this type have been developed and used to simulate bacterial colonies and biofilms \cite{rudge, storck,farrell_mechanically_2013, grant_2014, volfson_biomechanical_2008, PGhosh2015,head_pre}; they differ in their choices of which physical interactions to include (and how to include them), as well as in how they account for biological details  such as bacterial shape and metabolism. 

\begin{center} \noindent\rule{4cm}{0.4pt} \end{center}
\vspace{5mm}

\noindent

In the next three sections, \ref{sec:phasesofgrowth}, \ref{sec:buckling} and \ref{sec:clone}, we  discuss three examples of interesting phenomena produced by bacterial growth in spatially structured environments: fingering instabilities at the edges of colonies and biofilms, the transition from 2-dimensional to 3-dimensional colony growth and the emergence of genetically segregated sectors during expansion of a colony. In choosing these examples, we focus on phenomena that are unconnected with bacterial motility, since motility-induced collective phenomena in bacterial populations, and in general, have been extensively described elsewhere (see, e.g. \cite{cates_diffusive_2012,catestailleur,kearns,marchetti,cates_review2,vicsek,toner,gregoire}).

\subsection{Example: Fingering instabilities at the interfaces of bacterial colonies and biofilms}\label{sec:phasesofgrowth}

As we have already mentioned, the expanding edge of a growing bacterial colony, and the surface of a growing biofilm, are examples of growing interfaces. Depending on the growth conditions, the interface can be smooth, rough or feature long finger-like shapes (for colonies) or ``mushrooms'' (for biofilms) \cite{benjacob_review,benjacob_review2,bonachela_universality_2011,klausen,vicsek1990,ruzicka,wakita,ohgiwara}; see Fig. \ref{fig:interface}A-C. 

\begin{figure}
\centering
\includegraphics[clip, trim = 0cm 3cm 2cm 1cm, width=0.9\linewidth]{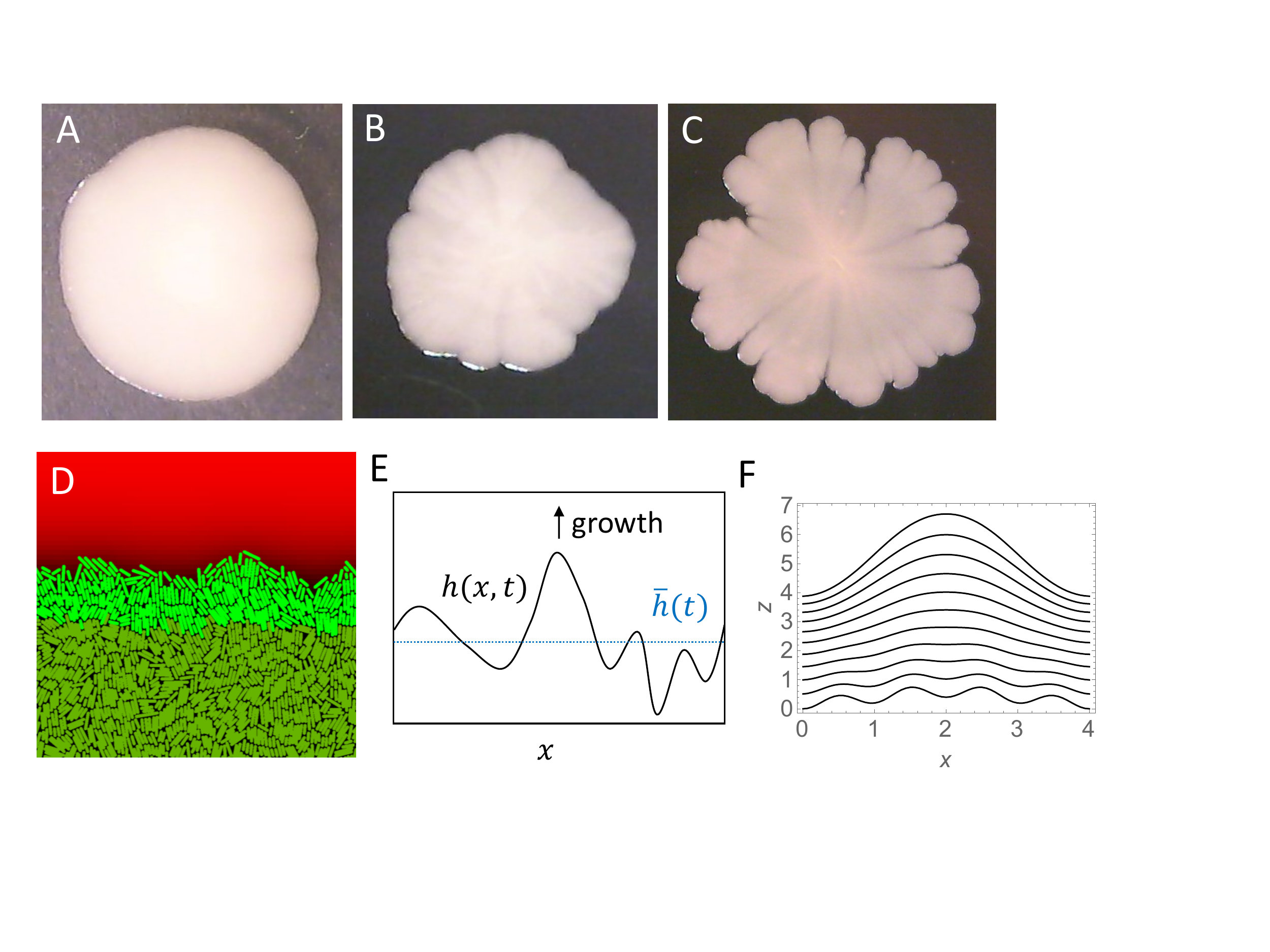}
\caption{\label{fig:interface}(A-C) Examples of {\it E. coli}  colonies (on 2mm-thick, 2\% agarose infused with LB) with different roughness of the colony boundary: smooth (A), rough (B), and branched (C). These different shapes have been obtained by using different strains (MG1655 for B, MG1655$\Delta$fimA$\Delta$fliF for A), or incubating colonies of MG1655 for different amounts of time at 37C (B = short time, C = long time). (D) Boundary of a simulated colony (simulation details as in Ref. \cite{farrell_mechanical_2017}). The nutrient concentration is shown as different shades of red (brightest colour = highest concentration). Replicating cells are shown in bright green, whereas stationary cells with no access to nutrients are shown in dark green. (E) Schematic illustration of the model from Eq. (\ref{eq:simplesurfgr}). (F) Interface profiles $h(x,t)$ obtained by numerically solving Eq. (\ref{eq:simplesurfgr}) for $L=4,\zeta=0.02,b=1,f(h)=1/(1+e^{-h})$, and $t=0,\dots,10$. The initial condition is a superposition of two sine functions with periods $4$ and $1$. The oscillation with period $1$ is damped ($1<\Lambda=2\pi\sqrt{\zeta f(0)/f'(0)}\approx 1.256$), whereas the one with period $4$ grows in time.}
\end{figure}

Various theoretical approaches have been used to model the shape of these interfaces, ranging from continuum equations \cite{klapper_finger_2002,giverso_emerging_2015} to individual-based models \cite{ali_reproduction-time_2012, farrell_mechanically_2013, farrell_mechanical_2017}. Rather than describing these in detail here, we will instead use a simple toy model to illustrate some basic factors that can affect the shape of the growing edge of a bacterial population. Although the model that we will present here is unrealistic in many ways, it has the advantage of allowing a  simple mathematical analysis. 

Let us imagine a two-dimensional population of bacteria which expands in the $z$ direction and is confined between two walls in the $x$-direction (Fig. \ref{fig:interface}E). This could represent a 2D bacterial colony or a 2D section of a biofilm; here we will refer to it as a biofilm. The interface of the growing biofilm has profile $z=h(x,t)$ at time $t$. We assume that bacteria grow only in a narrow zone of width $b$, close to the interface because nutrient does not penetrate far into the biofilm. {\black{This assumption is based on simulations like that shown in Fig. \ref{fig:interface}D, in which 
both bacteria and nutrients are modelled explicitly; here the bacteria shown in bright green, which are close to the nutrient, are able to replicate, while the bacteria shown in dark green, which are far from the nutrient, are not able to replicate.}} Although we do not model the nutrient concentration field explicitly here, we will assume that parts of the interface that protrude in the $z$ direction experience a higher nutrient concentration because they are closer to a nutrient source (this would be the case in a typical biofilm flow setup \cite{tolker}). Thus, the local growth rate depends on the height of the interface: $g(x,t) = f(h(x,t)-\bar{h}(t))$, where $\bar{h}(t)$ is the average height at time $t$ (Fig. \ref{fig:interface}E). We also suppose that the interface has a ``stiffness'', or a tendency to be flat. This is an {\em ad hoc} assumption, but it mimics, to some extent, adhesion between the bacteria. The dynamics of the interface can then be described approximately as
\bq
	\frac{\partial h}{\partial t} = bf(h(x,t)-\bar{h}(t))\left(\zeta \frac{\partial^2 h}{\partial x^2} + 1\right) . \label{eq:simplesurfgr}
\eq
In Eq. (\ref{eq:simplesurfgr}), the term $\zeta \partial^2 h/\partial x^2$ accounts for the surface stiffness by favouring growth in concave regions of the interface (troughs) and disfavouring growth in convex regions (peaks). 

To see how this model can produce interesting behaviour, let us  make a small perturbation around an initially flat interfacial profile: 
\bq
	h(x,t)=\bar{h}(t) + \epsilon(t) e^{i k x}.
\eq
Inserting this into Eq. (\ref{eq:simplesurfgr}) and expanding to first order in the magnitude of the perturbation, $\epsilon\ll 1$,  we obtain the following equation for  $\epsilon(t)$:
\bq
	\frac{d\epsilon(t)}{dt} = b\epsilon(t)\left [f'(0) - \zeta k^2f(0)\right],
\eq
where $f'(0)$ is the derivative of the growth function $f(h-\bar{h})$, evaluated for $h=\bar{h}$.  Thus, $\epsilon$ is predicted to grow in time for perturbations whose wavenumber $k$ obeys
\bq
	k^2 < f'(0)/(\zeta f(0)).
\eq
This condition is equivalent to stating that interfacial ``bumps'' of dimension  $\Lambda=2\pi/k$ will tend to grow if $\Lambda > 2\pi\sqrt{\zeta f(0)/f'(0)}$. This means that the interface will be unstable to the growth of finger-like protrusions, provided  the width of the system $L$ is large enough to allow such protrusions to develop, {\em i.e.} for systems of size $L > 2\pi\sqrt{\zeta f(0)/f'(0)}$ (Fig. \ref{fig:interface}F). Thus, a transition from a smooth to a fingered front is predicted to occur for a growing bacterial population if (i) the spatial extent $L$ of the population is big enough, (ii) the stiffness $\zeta$ of the interface is small enough and (iii) the growth function $f$ depends strongly enough on the height - {\em e.g.} due to rapid nutrient consumption or slow nutrient diffusion \cite{farrell_mechanically_2013}. 

This toy model is of course highly simplistic -- among other deficiencies, it does not account for the dynamics of the nutrient, or for changes in the thickness of the growing layer as the colony expands. Nevertheless it illustrates how instabilities can arise from the coupling between the shape of the growing colony/biofilm interface and the local availability of nutrient. Similar phenomena, driven by the same interface-nutrient coupling, also arise in more realistic models  \cite{dockery_2001,klapper_finger_2002,giverso_emerging_2015,kragh_2016,melaugh_2016}.

\subsection{Example: 2D to 3D transition in bacterial colony growth}\label{sec:buckling}

Another interesting feature of bacterial colony or biofilm growth is the transition from 2D to 3D growth. Starting from a single cell seeded on an agarose gel surface, a colony initially spreads as a 2D layer of cells on the surface, but it later develops into a 3D structure. If the agarose gel surface is covered by a glass coverslip (with the bacteria sandwiched between the agarose and the coverslip), then the colony becomes 3-dimensional by growing into the agarose layer. However, if there is no bounding coverslip, the colony instead expands into the space on top of the agarose layer. Biofilm growth on a solid surface also often starts with the proliferation of flat microcolonies, which later expand vertically. This 2D to 3D transition has parallels in the growth of some cancer tumours \cite{liotta} and in embryonic development \cite{dev_book}.

Experimental work on {\em E. coli} colonies growing on agarose suggests that mechanical forces are likely to play an important role in the 2D to 3D transition \cite{volfson_biomechanical_2008, grant_2014, su_bacterial_2012}. In a setup where bacteria are sandwiched between the agar and a glass coverslip, microscopic tracking of the growth of colonies from single cells reveals a well-defined ``buckling transition'' at which bacteria start to invade the agarose, leading eventually to 3D growth (Fig.   \ref{fig:2dto3d}A-C) \cite{grant_2014}. In this transition, the first cells to invade the agarose are usually located close to the centre of the 2D colony. Moreover, the average size of the 2D colony at the moment when this transition happens depends non-monotonically on the concentration (and hence the stiffness) of the agarose gel: it happens later ({\em i.e.} at larger colony area) for intermediate agarose stiffness. Using individual-based simulations, Grant {\em et al.} \cite{grant_2014} could match these experimental results, under the assumption that the friction coefficient between the bacteria and the agarose has a particular non-linear dependence on the agarose stiffness. 

\begin{figure}
\centering
\includegraphics[clip, trim = 0cm 1cm 0cm 0cm, width=0.85\linewidth]{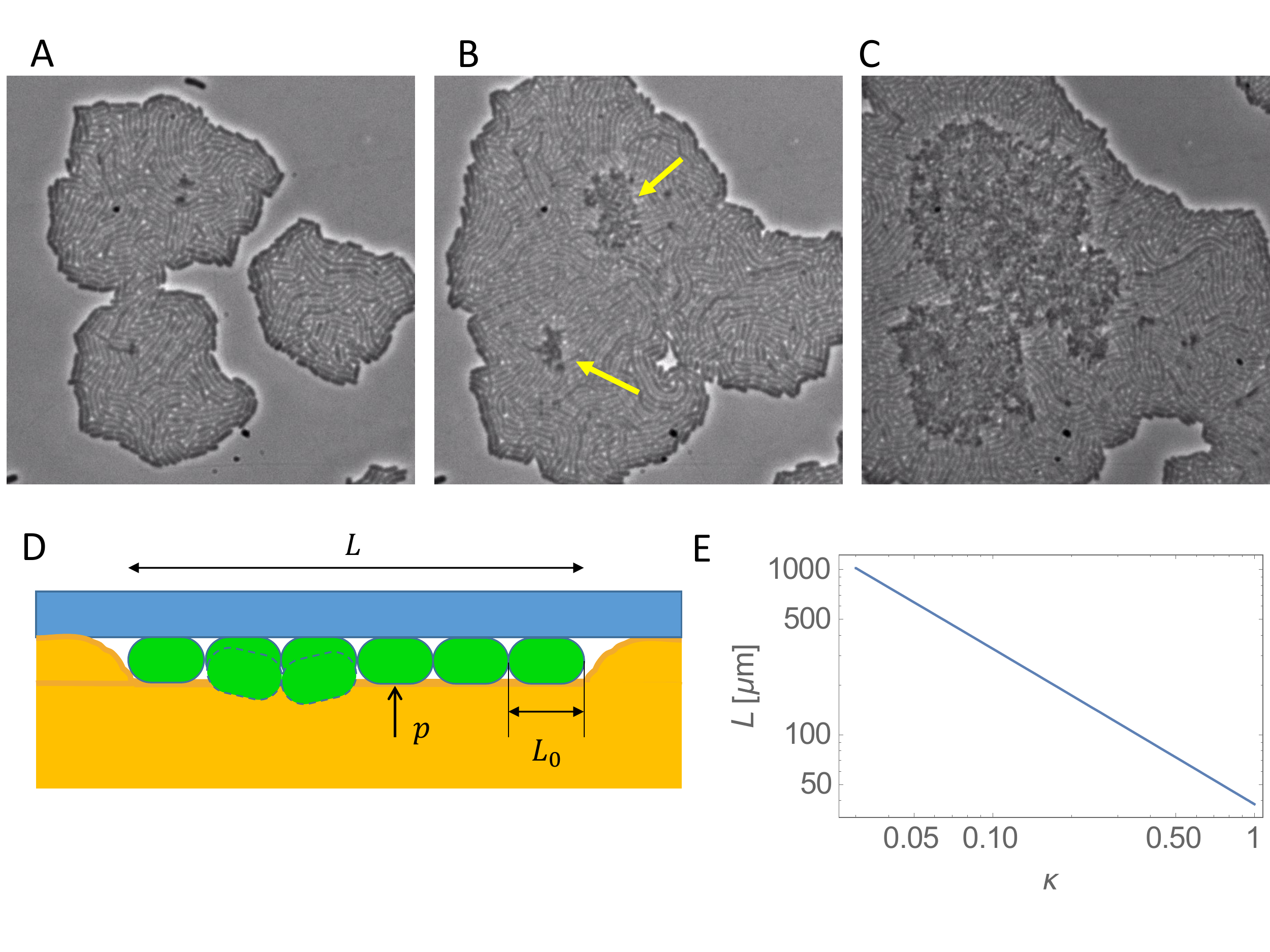}
\caption{\label{fig:2dto3d}(A-C) 2D to 3D transition in bacterial colonies. The arrows indicate locations where the colony has invaded the agarose and a second layer of cells has begun to form. (A) An image taken just before the transition. (B) An image of the same colony taken just after the transition. (C) An image taken when the second layer of cells is already well-developed. (D) A simple model of the ``buckling'' transition, for a 1D chain of bacteria (see main text). (E) Length of the bacterial chain at the onset of the buckling transition, as a function of the friction coefficient $\kappa$. The parameters are $g=2$h$^{-1},L_0=1\mu$m, $b=10^{7}$Pa$^{-1},w_0=10^{-7}$h$^{-1},p=10^5$Pa.}
\end{figure}

While Grant {\em et al.}'s simulations were quite complex, the basic physics that may control the invasion transition can be illustrated with a much simpler model (Fig. \ref{fig:2dto3d}D). Let us imagine a 1D chain of bacteria, extending from $x=-L/2$ to $x=L/2$. The bacteria elongate at rate $g$ and so the chain length $L(t) = L_0\exp(g t)$ increases with time. As the bacteria grow, they exert outward pushing forces on each other and experience inward forces due to friction with the surrounding medium (here assumed to be agarose). This produces a local stress $\sigma(x,t)$ within the chain. Because the frictional forces are transmitted along the chain of bacteria, we expect $\sigma(x,t)$ to be largest at the centre of the chain, $x=0$, and to increase in time as the chain elongates. This stress may cause the chain to buckle (Fig. \ref{fig:2dto3d}D); we denote the stress-dependent rate at which a bacterium buckles as $w(\sigma)$. We suppose that $w$ is small for small stress $\sigma$, but increases strongly for large $\sigma$. One form of $w(\sigma)$ consistent with this expectation is $w(\sigma) = w_0\exp{[b\sigma]}$, where $b$ is some constant; for illustrative purposes we will assume this form here (although it is not motivated by any mechanistic understanding of the buckling process).

Focusing on a particular position $x$ along the chain, the probability that the chain has not buckled at this position by time $t$ is $	\exp\left[-\int_0^t w(\sigma(x,t')) dt'\right]$, and the probability that the chain has not buckled at any position by time $t$ is $\exp\left[-\int_0^t \int_{-L(t')/2}^{L(t')/2} w(\sigma(x,t')) dx dt'\right]$. Therefore,  the probability $P(t)$ that a buckling event {\em has} happened by time $t$ is
\bq
P(t) = 1-	\exp\left[-\int_0^t \int_{-L(t')/2}^{L(t')/2} w(\sigma(x,t')) dx dt'\right].
\eq
Now let us assume a particular form for the stress function: $\sigma(x,t) = k [L(t)/2-x]$ for $x>0$ and $\sigma(x,t) = k [L(t)/2+x]$ for $x<0$. This simply describes a linear decrease in stress from the centre to the edge of the chain, with the constant $k$ being related to the friction coefficient. This form of $\sigma$ might be expected if the frictional force generated by bacterial motion is equal for all bacteria, and the contributions of each bacterium sum up along the chain. 

Changing variables to $y=L/2-x$, using the symmetry of $\sigma(x,t)$ about $x=0$ and substituting in the chosen form of $w$, we obtain
\begin{eqnarray}
P(t) &=& 1-	\exp\left[-2\int_0^t \int_{0}^{L(t')/2} w(ky) dy dt'\right] \\\nonumber & =& 1-	\exp\left[-\frac{2w_0}{bk}\int_0^t \left(\exp{\left[bkL(t')/2\right]}-1 \right)dt'\right] .
\end{eqnarray}
We now use the fact that $L = L_0\exp{(g t)}$ to replace time $t$ by the chain length $L$. The probability $Q(L)$ that the chain has not buckled by the time it reaches length $L$ is
\begin{eqnarray} \label{eq:qofL}
Q(L) & =& 1-	\exp\left[-\frac{2w_0}{bkg}\int_0^{L} \left(\frac{\exp{(bkl/2)}-1}{l}\right)dl\right] \\\nonumber &=&
 1-	\exp{\left[-\frac{2w_0}{bkg}\left[-\gamma + {\rm{Shi}}{\left[\frac{bkL}{2}\right]} + {\rm{Chi}}\left[\frac{bkL}{2}\right] - \log{\left[\frac{bkL}{2}\right]}\right]\right]},
\end{eqnarray}
where $\gamma$ is the Euler-Mascheroni constant and Shi and Chi are the sinh and cosh integral functions. The dependence of the probability $Q(L)$ on the length of the bacterial chain $L$ arises only from terms in the combination $bkL$. This leads to our first important observation: the critical size at which the buckling transition happens is expected to scale approximately as $1/(bk)$:  {\em i.e.} it decreases with increasing friction/adhesion $k$ and increasing growth rate $b$. The most likely length $L$ of the chain  at which the transition happens can be determined from
\bq
  \frac{2w_0}{bkg}\left[-\gamma + {\rm{Shi}}{\left[\frac{bkL}{2}\right]} + {\rm{Chi}}\left[\frac{bkL}{2}\right] - \log{\left[\frac{bkL}{2}\right]}\right] \approx 1.
\eq

 To relate the predictions of this simple model to experimental results such as those of Grant {\em et al.}, the coefficient $k$ in the model must be related  to the friction coefficient $\kappa$ between bacteria and the agarose / glass surfaces, and the stress $p$ that pushes a bacterium against the glass surface, due to elastic compression of the agarose (Fig. \ref{fig:2dto3d}D). Dimensional analysis suggest that $k=\kappa p/L_0$. Fig. \ref{fig:2dto3d}E shows the resulting predictions of this simple model for the chain length at the onset of buckling, as a function of $\kappa$, for $g=2$h$^{-1},L_0=1\mu$m, and with the parameters $b=10^{7}$Pa$^{-1}$ and $w_0=10^{-7}$h$^{-1}$ chosen so that the result is comparable with the experimentally observed buckling diameter ($\approx 50\mu$m) of a 2D colony for  $\kappa=0.7$ and $p=10^5$Pa \cite{grant_2014}. The model predicts that the size of the colony upon buckling decreases with the friction coefficient $\kappa$ which characterises the strength of cell-agarose interactions.
One can also use similar arguments to show that the average position at which the chain buckles is very close to its centre, in agreement with the experimental results  \cite{grant_2014}.

This model, while it is undoubtedly simplistic, provides some insight into the effects of friction on the 2D to 3D transition.  
 More generally, understanding the 2D to 3D transition in bacterial colonies presents a host of interesting challenges. These include analysing simple statistical physics models like the one described here, developing individual-based simulations that explicitly include interactions with the surrounding elastic medium, and carrying out experimental measurements of the frictional and adhesion forces between bacteria and agarose and glass surfaces. It is also important to note that the buckling transition that we have discussed here is only the first stage the development of a 3D bacterial colony. Once buckling has happened, the subsequent development from a two-layered structure to a larger 3D colony also  presents beautiful and interesting phenomena which remain to be explained \cite{su_bacterial_2012}.

\subsection{Example: Formation of clonal sectors in growing bacterial colonies}\label{sec:clone} 

\begin{figure}
\centering
\includegraphics[clip, trim = 0cm 7cm 0cm 0cm, width=0.85\linewidth]{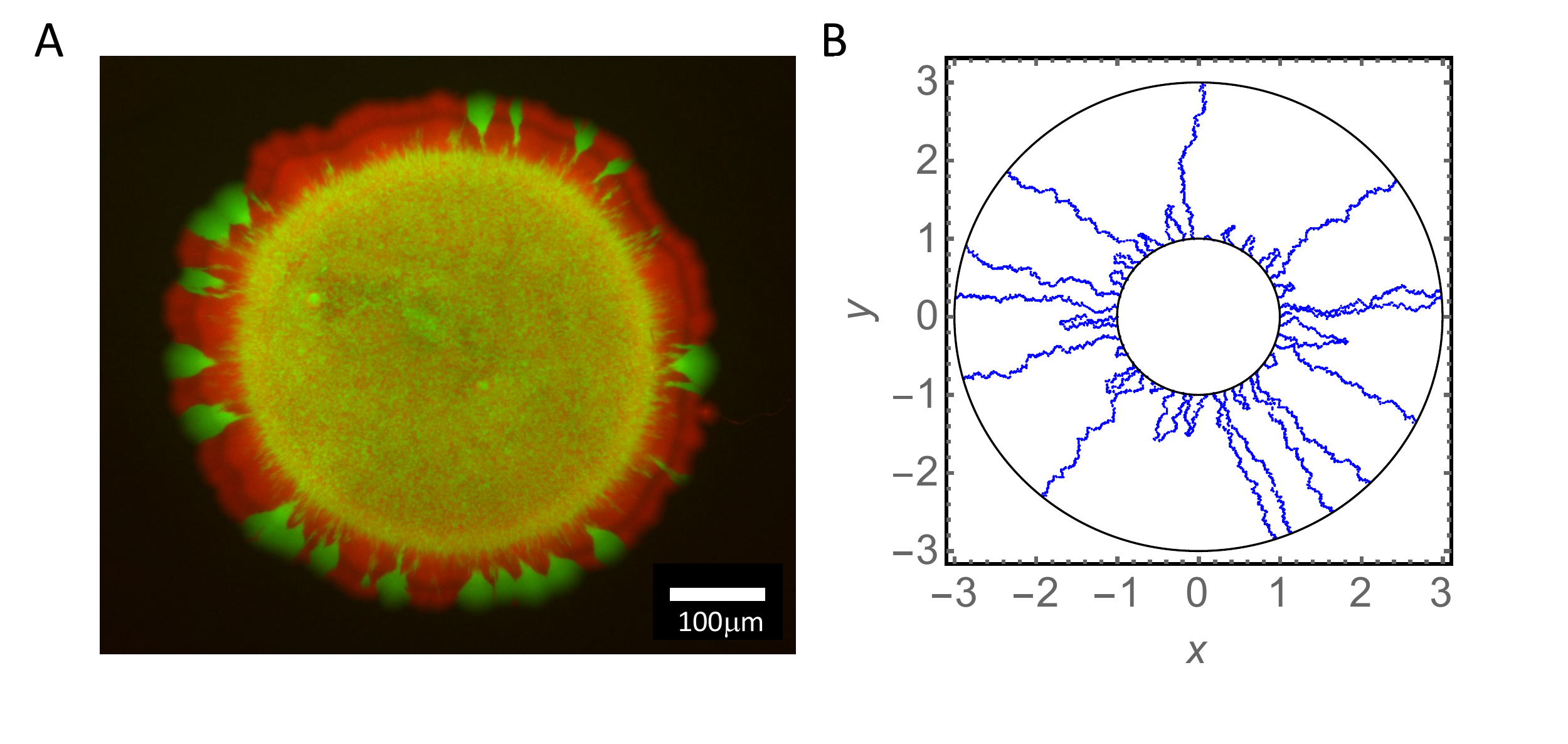}
\caption{(A) An expanding population of fluorescently-labelled {\em E. coli} cells, growing on the surface of a nutrient agar pad. The population was initiated from a drop containing a 50:50 mixture of cells labelled in two different colours (here shown red and green). The population is mixed (appears yellow) in the region of the initial drop, but has segregated into clonal sectors at its expanding edge. Image courtesy of Diarmuid Lloyd. (B) Results of a simulation in which sector boundaries are modelled as annihilating random walks, as described by Eq. (\ref{eq:langw}) with $D=0.02, R_0=1, R=3$. The simulation starts with $50$ random walkers.}
\label{fig:sectors}
\end{figure}

A very interesting feature of the growth of bacterial colonies and biofilms is the spatial distribution of lineages within the population -- or in other words, the locations of the descendants of a particular ``founder cell''. Fig. \ref{fig:sectors}A shows the outcome of a simple experiment in which a bacterial colony is initiated not from a single cell, but from a droplet containing a mixture of two strains of {\em E. coli} which are identical except that they produce different-coloured fluorescent proteins (here shown red and green). The area covered by the initial droplet appears yellow, indicating a mixture of red and green cells. In the surrounding regions, however, where the population has expanded out from the initial droplet, a striking pattern of red and green sectors is visible. This implies genetic segregation: the descendants of different cells within the founder population occupy different regions of space \cite{hallatschek_2007}. The same phenomenon occurs for other microorganisms \cite{hallatschek_2010,foster2011}, and for colony growth in different geometries \cite{hallatschek_2010,korolev_2012}. 


The emergence of sectors is closely connected with the fact that (after an initial period of exponential growth) only bacteria that are close to the expanding edge of the colony are able to replicate; deeper in the colony nutrient becomes depleted and waste products may accumulate. Demographic fluctuations at the growing colony front can cause a bacterial lineage to become ``trapped'' behind the front, in which case it cannot proliferate further. Thus, a stochastic process is at play, in which some lineages come to dominate the growing front ({\em{i.e.}} form sectors) while others are buried behind the front. 

To better understand this process, we follow Ref. \cite{hallatschek_2010} and imagine that the growing layer is infinitely thin and circular symmetric, such that the proliferating bacteria are located on the perimeter of a circle of radius $R=vt$ expanding with constant velocity $v$. Let us suppose that the initial radius of the circle ({\em i.e.} the radius of the drop of bacteria that is deposited on the agar) is $R_0$. We divide the perimeter into sectors, and we track the positions of the sector boundaries on the perimeter of the circle {\footnote{In this calculation we do not specify the number of bacteria in each sector as this turns out not to be important.}.
As time goes on, the bacteria within each sector proliferate, or become lost behind the growing front, in a stochastic process. While the size of a sector will increase on average as the colony expands, at any given moment in time it may fluctuate either upwards or downwards. 
A sector may even contract so much that it vanishes altogether, representing loss of the lineages of the bacteria within that sector.  This stochastic dynamics can be modelled by the following Langevin equation for the arc length $w$ occupied by a sector:
\bq
	\frac{{\rm d}w}{{\rm d}R} = \frac{w}{R} + \sqrt{4D} \eta(R). \label{eq:langw}
\eq
with the initial condition $w(R_0)=w_0$. Note that because the colony radius $R$ increases linearly with $t$, tracking the dynamics as a function of $R$ is equivalent to tracking it as a function of time $t$. 
In Eq. (\ref{eq:langw}), the first term on the right hand side accounts for the radial expansion of the colony, which ``stretches'' the sector. The second term accounts for stochasticity in the replication events and local movements of bacteria  at the front; here $D$ is an effective diffusion constant and $\eta(R)$ represents uncorrelated Gaussian noise with zero mean and unit variance. \black{The lack of scaling of $\eta(R)$ with $R$ is due to inter-sector competition assumed to occur only at sector boundaries of constant width. 
We stress that Eq. (\ref{eq:langw}) defines an idealized mathematical model of a circular-symmetric, infinitesimally-thin edged colony; sector dynamics in real colonies may deviate from it due to edge roughness (see Secs. \ref{sec:ibm}, \ref{sec:phasesofgrowth}, and Ref. \cite{hallatschek_pnas}). } 
In the absence of the noise term,  Eq. (\ref{eq:langw}) predicts that $w$ increases deterministically as $w(R)=w_0 R/R_0$. For $D>0$, however, the arc length $w$ follows a biased random walk with a time-dependent diffusion constant. This is illustrated in Fig. (\ref{fig:sectors})B in which we plot trajectories of sector boundaries, simulated using Eq. (\ref{eq:langw}).

To proceed further, we introduce the angular size of the sector, $\phi=w/R$. In this coordinate, Eq. (\ref{eq:langw}) becomes
\bq
	\frac{{\rm d}\phi}{{\rm d}R} = \frac{\sqrt{4D}}{R} \eta(R). \label{eq:langp}
\eq
Thus, we see that the magnitude of the angular fluctuations decreases as the colony radius increases. Eq. (\ref{eq:langp}) can be translated into a Fokker-Planck equation \cite{gardiner_stochastic_2009}:
\bq
	\frac{\partial P(\phi,R|\phi_0,R_0)}{\partial R} = \frac{2D}{R^2} \frac{\partial^2 P(\phi,R|\phi_0,R_0)}{\partial \phi^2},  \label{eq:ex3}
\eq
where $P(\phi,R|\phi_0,R_0)$ is the probability that a sector has angular size $\phi$ when the colony radius is $R$, given that its size was $\phi_0$ at $R_0$. If a sector shrinks to angular size $\phi=0$ then we assume it cannot recover (since the lineage becomes lost behind the growing layer): this implies the boundary condition $P(0,R|\phi_0,R_0)=0$. We also set $P(\infty,R|\phi_0,R_0)=0$ because sectors cannot become arbitrarily large\footnote{Actually, $\phi$ cannot be larger than $2\pi$ but we expect most sectors to be much smaller than this if the initial number of sectors is large. Assuming an absorbing boundary at $\phi\to\infty$ simplifies  the calculations.}.
With these boundary conditions the solution of Eq. (\ref{eq:ex3}) is
\bq
	P(\phi,R|\phi_0,R_0) = \frac{1}{\sqrt{2 \pi \sigma^2}} e^{-\frac{(\phi +\phi_0)^2}{2 \sigma^2}} \left(e^{\frac{2 \phi \phi_0}{\sigma^2}}-1\right),  \label{eq:ppp}
\eq
where $\sigma^2(R)= 4D(R_0^{-1} - R^{-1})$. This result can be obtained using the method of images and taking a Fourier transform of Eq. (\ref{eq:ex3}) \cite{rednerbook}. For small initial sector size $\phi_0\ll 2\pi$ we can expand Eq. (\ref{eq:ppp}) to first order in $\phi_0$,
\bq
	P(\phi,R|\phi_0,R_0) \cong \sqrt{2/\pi} \frac{\phi\phi_0}{\sigma^3} \exp\left(-\frac{\phi^2}{2\sigma^2}\right). \label{eq:ppp1}
\eq
From an experimental point of view, one can easily measure the sizes of sectors in relatively large colonies (e.g. $R \sim $ a few mm), but it is much harder to measure sectors in very small colonies. Therefore we would like to use Eq. (\ref{eq:ppp1}) to predict the distribution of sizes of surviving sectors, in the large colony limit $R \to \infty$. We first normalize Eq. (\ref{eq:ppp1}) to obtain the distribution $P_{\rm surv}(\phi,R|\phi_0,R_0)$ of sector sizes, conditioned on sector survival:
\bq
	P_{\rm surv}(\phi,R|\phi_0,R_0) \cong \frac{\phi}{\sigma^2} \exp\left(-\frac{\phi^2}{2\sigma^2}\right).
\eq
The mean angular sector size in the limit $R\to\infty$ is thus
\ba
	\left<\phi(R\to\infty)\right> &=& \int_0^\infty \phi P_{\rm surv}(\phi,R\to \infty|\phi_0,R_0) d\phi = \sqrt{\frac{\pi\sigma^2(R\to\infty)}{2}} \nonumber \\
    &=& \sqrt{\frac{2\pi D}{R_0}},
\ea
where we have used $\sigma^2(R\to\infty) = 4D/R_0$. The average number of sectors is thus 
\bq
	N_{\rm sectors}(R\to\infty) = \frac{2\pi}{\left<\phi(R\to\infty)\right>} = \sqrt{\frac{2\pi R_0}{D}}.
\eq
Interestingly, this theory predicts that the number of sectors in the large colony limit is finite, showing that coexistence between different lineages is possible.
Note that $N_{\rm sectors}(R\to\infty)$ is independent of the initial number of sectors, as long as this initial number is large (small $\phi_0$), but it depends on the initial radius $R_0$ of the colony. Individual-based simulations of colony growth and experiments with bacteria growing on agar plates confirm this prediction  \cite{ali_reproduction-time_2012,rudge, gralka2016, farrell_mechanical_2017} and show that it remains qualitatively true if the growing layer has finite thickness. Simulations and extensions of the theory can also be used to predict what happens when the colony contains mixtures of bacteria with different growth rates \cite{hallatschek_2010,gralka2016}, when the bacteria are able to undergo horizontal gene transfer between neighbouring cells  \cite{kuba+BW,freese}, or when the growing population encounters obstacles  \cite{moebius2016}.

This theoretical analysis provides an example of the use of statistical physics to understand a complex biological phenomenon. However it does not explain what features of the growth process control the diffusion constant $D$, which plays a critical role in determining the number of sectors. Indeed, the number of sectors has been observed to differ between different organisms: fewer sectors are observed for the yeast {\em S. cerevisiae} than for {\em E. coli}, and also, intriguingly, fewer sectors are observed for a spherical mutant of {\em E. coli} than for the usual rod-shaped  {\em E. coli}  cells \cite{hallatschek_2007,hallatschek_2010}. Individual-based simulations have an important role to play in explaining these observations; these simulations have already pointed to mechanical interactions between cells and the surface on which they grow as major players in determining $D$ \cite{farrell_mechanical_2017} .

Genetic segregation within an expanding bacterial population, as described in this example, has important evolutionary implications, since it significantly affects the ``surfing probability'', or the probability that a mutant arising at the front of an expanding population forms a macroscopic sector \cite{gralka2016}. This is relevant, for example, to the evolution of antibiotic resistance in bacterial biofilms. A similar problem arises in the evolution of drug resistance in cancer tumours \cite{waclaw_spatial_2015}.

\section{Conclusions and outlook}

The primary purpose of this review has been to illustrate the rich array of beautiful and interesting phenomena displayed by growing bacterial populations. These phenomena are intrinsically non-equilibrium and many of them lend themselves naturally to  analysis using the tools of statistical physics.

Research at the interface between microbiology and statistical physics can have great benefits for both fields. Statistical physics models can cut through biological detail and provide insight into basic biological mechanisms, when they are properly constructed with knowledge of the underlying biology. The application of statistical physics to biological problems can also generate new non-equilibrium models that drive further development in statistical physics. Fruitful interplay between statistical physics and biology is nothing new:  examples include  the totally asymmetric exclusion process \cite{blythe_review}, which was introduced  as a model for cellular protein production \cite{mcdonald} and has since become a paradigm for non-equilibrium transport processes. What we aim to highlight here is the attractiveness of bacterial populations, specifically, as subjects for statistical physics models. We believe that this is a timely topic, from the point of view of both physics and biology. From a physics point of view, new non-equilibrium physics is emerging from the study of active systems, which have up to now been mainly focused on motile particles (``swimmers'') {\black{\cite{cates_diffusive_2012,catestailleur,najafi,golestanian2007,pooley}}}. Yet bacteria are also active in many other ways: they grow, divide, secrete signals and macromolecules, and interact chemically and mechanically in a complex way. The statistical physics of these behaviours, especially growth in dense assemblies of bacteria, has just started to be explored.

From a biological point of view, the growing threat of antimicrobial resistance, increasing awareness of the role of  biofilms in infection, and the growing understanding of the importance of microbes in gut health have raised the profile of microbiology in recent years. There has also been a resurgence in interest among microbiologists in both fundamental growth phenomena and in the use of mathematical models to explain them. Thus, statistical physics models of bacterial growth have the potential to make a significant impact.

We would also like to highlight here the importance of experiments. Many (although not all) of the bacterial growth phenomena discussed in this review arise in rather simple microbiological experiments. Our own experience is that even a brief immersion into experimental work with bacteria  can greatly improve one's ability to develop relevant, realistic and interesting statistical physics models. Moreover this is often a fun experience! We therefore advocate spending some time in the lab to even the most hardened theoretician, if it is at all possible.

\ack
The authors thank Lucas Black,  Richard Blythe, Rebecca Brouwers, Timothy Bush, Martin Carballo Pacheco, Mike Cates, Luca Ciandrini, Pietro Cicuta, Steven Court, Susana Direito, Martin Evans, Andrew Free, Philip Greulich, Oskar Hallatschek, Bhavin Khatri, Elin Lilja, Diarmuid Lloyd, Cat Mills, Nikola Ojkic, Eulyn Pagaling, Jakub Pastuszak, Wilson Poon, Patrick Sinclair, Sharareh Tavaddod, Daniel Taylor, Simon Titmuss, Juan Venegas-Ortiz, Paolo Visco, Patrick Warren and Ellen Young for many discussions. Special thanks are due to Bhavin Khatri, Joost Teixeira
de Mattos, Alex ter Beek, Martijn Bekker, Tanneke den Blaauwen, Yasuhiko Irie and Diarmuid Lloyd, who provided original data or images for the figures shown here.  
RJA was supported by a Royal Society University Research Fellowship, by the Human Frontier Science Program under grant RGY0081/2012, by the US Army Research Office under grant 64052-MA, by EPSRC under grant EP/J007404/1, and by ERC Consolidator Grant 682237-`EVOSTRUC'. BW was funded by a Leverhulme Trust Early Career Research Fellowship and by a Royal Society of Edinburgh Research Fellowship.

\section*{References}


\end{document}